\documentclass[10pt]{article}
\usepackage{yfonts}
\usepackage{epsfig}
\usepackage{subfigure}
\usepackage{supertabular}
\usepackage{amsmath}
\usepackage{amssymb}
\usepackage{bbold}
\usepackage{cite}
\usepackage[sans]{dsfont}
\usepackage{graphicx}
\usepackage[dvips]{color}
\newcommand{\cleqn}{\setcounter{equation}{0}}
\newcommand{\newc}{\newcommand}
\newc{\om}{\omega}
\newc{\sig}{\sigma}
\newc{\beq}{\begin{equation}}
\newc{\eeq}{\end{equation}}
\newc{\beqn}{\begin{eqnarray}}
\newc{\eeqn}{\end{eqnarray}}
\newc{\bsym}{\boldsymbol}
\newc{\ol}{\overline}
\def\bvec#1{\raise1.5ex\hbox{$\rightarrow$}\mkern-16.5mu #1}

\def\m#1{\mathcal#1}

\begin{document}

\title{\hfill ~\\[-30mm]
       \hfill\mbox{\small UFIFT-HEP-08-14}\\[30mm]
       \textbf{The Flavor Group $\bsym{\Delta(6n^2)}$}}
\date{}
\author{\\
        J. A. Escobar,\footnote{E-mail: {\tt jescobar@phys.ufl.edu}}~~
        Christoph Luhn\footnote{E-mail: {\tt luhn@phys.ufl.edu}}~~
       \\ \\
  \emph{\small{}Institute for Fundamental Theory, Department of Physics,}\\
  \emph{\small University of Florida, Gainesville, FL 32611, USA}}
\maketitle

\begin{abstract}
\noindent Many non-Abelian finite subgroups of $SU(3)$ have been used
to explain the flavor structure of the Standard Model. In order
to systematize and classify successful models, a detailed
knowledge of their
mathematical structure is necessary. 
In this paper we shall therefore look closely at the series of finite
non-Abelian groups known as $\Delta(6n^2)$, its smallest members
being $\mathcal S_3$ ($n=1$) and $\mathcal S_4$ ($n=2$). For arbitrary $n$, we
determine the conjugacy classes, the irreducible representations, the
Kronecker products as well as the Clebsch-Gordan coefficients.
\end{abstract}

\vfill
\newpage
\section{Introduction}
\cleqn

The experimental observation that neutrinos have non-zero mass has opened up a
window to physics beyond the Standard Model. Advances in the setup of several
neutrino experiments furthermore revealed that the mixing in the lepton
sector features two large and one small angle. Its numerical values lead to a
Maki-Nakagawa-Sakata-Pontecorvo~(MNSP) matrix which is intriguingly close to
the so-called tri-bimaximal mixing pattern~\cite{tribi}
\begin{eqnarray}
\mathcal U_{\mathrm{tri-bi}} &=&
\begin{pmatrix} 
\sqrt{\frac{2}{3}} & \frac{1}{\sqrt 3} & 0 \\
-\frac{1}{\sqrt 6} &\frac{1}{\sqrt 3} &-\frac{1}{\sqrt 2} \\
-\frac{1}{\sqrt 6} &\frac{1}{\sqrt 3} &\frac{1}{\sqrt 2} 
 \end{pmatrix} \ .
\end{eqnarray}
The most promising explanation for this remarkable fact seems to be the
existence of an underlying discrete
symmetry~\cite{S3,A4,S4,delta27,binaryA4,Q2n,D5,psl,seidl,other,overview}. Having
chosen a preferred non-Abelian finite symmetry group~\cite{finitesubgroup},
the three generations of fermions as well as the (extended) Higgs sector
particles are assigned to irreducible representations of the group. The theory
can then be formulated by  writing down all couplings which are allowed by the
symmetry group. A good understanding of the group structure is therefore an
essential first step in our endeavor to explain the physics of flavor.

The mathematical details of the groups of type $\Delta(3n^2)$ were worked out
in Ref.~\cite{LNR}. It is the purpose of this article to give an analogous
account for the groups of type $\Delta(6n^2)$. These two series of groups are
particularly interesting since their smallest members $\Delta(12) = \mathcal
A_4$~\cite{A4} and $\Delta(27)$~\cite{delta27}, as well as $\Delta(6) =
\mathcal S_3$~\cite{S3} and $\Delta(24) = \mathcal S_4$~\cite{S4} have been
applied successfully in numerous models of flavor. 
Discussing these finite subgroups of $SU(3)$ for arbitrary values of $n$ 
might be helpful in constructing  more constrained 
models with larger symmetry groups. We also hope that our general presentation
will serve as a tool for a systematic investigation of these flavor groups.

In Section~\ref{sclass}, we first determine the conjugacy classes of the groups
$\Delta(6n^2)$. All irreducible representations are obtained from the
six-dimensional induced representations in Section~\ref{sirreps}. A possible
choice for labeling the six-dimensional irreps is shown in
Appendix~\ref{standard}. The Kronecker products, presented in
Section~\ref{product}, can then be calculated from the character 
table. Appendix~\ref{calcdetail} details two ways of obtaining them, including
the explicit construction of the Clebsch-Gordan coefficients. We conclude in
Section~\ref{outlook}.


\section{The Structure of $\bsym{\Delta(6n^2)}$}\label{sclass}
\cleqn
The group $\Delta(6n^2)$ is a non-Abelian finite subgroup of $SU(3)$ 
of order $6n^2$. It is isomorphic to the semidirect product of the 
$\m{S}_3$, the smallest non-Abelian finite group, with
$(\m{Z}_n \times \m{Z}_n)$\cite{Bovier:1980gc},

$$\Delta(6n^2)~\sim~(\m{Z}_n \times \m{Z}_n)\rtimes \m S_3\ .$$
With the above in mind we now give the presentation of
$\Delta(6n^2)$: 

\begin{equation}\label{eq:pres1}
a^3 ~=~ b^2 ~=~ (ab)^2 ~=~ c^n ~=~ d^n ~=~1, 
\end{equation}
\begin{equation}
cd ~=~ dc ,
\end{equation}
\begin{equation}\label{eq:pres2}
\begin{array}{cccccc}
a c a^{-1}  &~=~&  c^{-1} d^{-1}, \quad
&a d a^{-1} &~=~&  c,             \\ 
b c b^{-1}  &~=~&  d^{-1},        
&b d b^{-1} &~=~&  c^{-1}.      \\    
\end{array}
\end{equation}
The elements $a$ and $b$ are the generators of the $S_3$ while
$c$ and $d$ generate $(\m{Z}_n \times \m{Z}_n)$.
The last two lines are consequences of the semidirect product.

Note that when given the generators of the two
groups involved in a semidirect product it is usually 
possible to obtain completely different sets of 
consistent group conjugations. Each separate set correspond to a 
possible distinct group.

Finally we should mention that in general any element of the group, which we will make use of occasionally, can be written in the form

\begin{equation} \label{eq:genform1}
 g ~=~ a^{\alpha} b^{\beta} c^{\gamma} d^{\delta} 
 \ ,
\end{equation}
where $\alpha$, $\beta$, $\gamma$, $\delta$ are integers. 
\subsection{Conjugacy Classes}
Our next step is to determine the conjugacy classes of the
group. Keeping in mind that the first class is always the
identity class, $1C_1(e) = \{e\}$. To obtain the remaining
classes we first need some classification principle, the 
following set of conjugations will give us such a principle.

\begin{equation}
\begin{array}{cccccc}
cac^{-1} &=& ac^{-1}d, \qquad &dad^{-1} &=& ac^{-1}d^{-2}\ ,\\
cbc^{-1} &=& bc^{-1}d^{-1},   &dbd^{-1} &=& bc^{-1}d^{-1}\ ,\\
bab^{-1} &=& a^2 ,            &aba^{-1} &=& a^2 b\ .
\label{action1}
\end{array}
\end{equation}
The above shows that conjugation by the Abelian
elements $c$ and $d$ on $a$ and $b$ will not alter their
powers. The last line, along with Eq.
(\ref{eq:pres1})-(\ref{eq:pres2}), is significant in that it places
elements of solely one and two powers of $a$ in the same type
of classes, and
those of single powers in $b$ with elements of $a^2 b$ and
$a b$ in another. 

We now begin to find the elements in each class, starting
with the classes formed by elements with no powers of either $a$ nor
$b$.

\begin{itemize}
\item{{\bf{Elements with neither $\boldsymbol{a}$} nor $\boldsymbol{b}$}}

We begin with the classes whose elements are of the form
$c^{\rho} d^{\sigma}$, where $\rho, \sigma= 0,1,...,n-1$.
We obtain the elements of this class by action of

\begin{equation} \label{eq:conjugate1a}
  g \left ( c^{\rho} d^{\sigma} \right )g^{-1},
\end{equation}
where $g=a^{\alpha}b^{\beta}$, $\alpha =0,1,2$, and $\beta=0,1$ . The result is the following 
class of six elements

\begin{equation} \label{eq:class1a}
\left \{   
c^{\rho}d^{\sigma}, c^{\sigma-\rho}d^{-\rho},
c^{-\sigma}d^{\rho-\sigma},
c^{-\sigma}d^{-\rho}, 
c^{\sigma-\rho}d^{\sigma},
c^{\rho}d^{\rho-\sigma} 
\right \}
\ ,
\end{equation}
the second and third obtained from conjugation with $a$ on
the first term, and the last three obtained by conjugating 
the first term with $b$ and then repeated
application of $a$. 

There are possible values for $\rho$ and $\sigma$ in which
the above class may collapse to a smaller set of elements,
providing, in effect, new classes. This collapse happens when 
any of the following conditions are met

\begin{equation}\label{eq:cond1}
  \rho+\sigma~=~ 0 ~\mathrm{mod} (n)\ , \quad 
  2\rho-\sigma~=~ 0 ~\mathrm{mod} (n)\ , \quad
  \rho-2\sigma~=~ 0 ~\mathrm{mod} (n)\ 
\ .
\end{equation}
The above can be obtained by equating the first term to the
fourth, the fifth, and then the sixth term.  
Each yields the same  class of three elements of the form 

\begin{equation} \label{eq:class2a}
\left \{   
c^{\rho}d^{-\rho}, c^{-2\rho}d^{-\rho},
c^{\rho}d^{2\rho}
\right \}
\ ,
\end{equation}
where $\rho= 1,2,...,n-1$. Closer inspection shows that the 
above can further collapse when 

\begin{equation}\label{eq:class3a}
  3\rho ~=~ 0~\mathrm{mod} (n)
  \ ,
\end{equation}
or in other words if $n = 3\mathbb{Z}$, leading to two
possibilities.

\begin{itemize}

\item[$(i)$]    $n\neq 3\,\mathbb{Z}$. Here we have only two
types of classes, those with six elements and those with three.

\begin{eqnarray}\label{eq:class4a}
n-1 & :&  3C_{1}^{(\rho)} ~=~
\left \{   
c^{\rho}d^{-\rho}, c^{-2\rho}d^{-\rho},
c^{\rho}d^{2\rho}
\right \},
~~~~ \rho=1,2,...,n-1, 
\\
\dfrac{n^2-3n+2}{6}& :& 6C_{1}^{(\rho,\sigma)}~=~
\left \{   
c^{\rho}d^{\sigma}, c^{\sigma-\rho}d^{-\rho},
c^{-\sigma}d^{\rho-\sigma},
c^{-\sigma}d^{-\rho}, 
c^{\sigma-\rho}d^{\sigma},
c^{\rho}d^{\rho-\sigma} 
\right \},
\end{eqnarray}
where $\rho,\sigma =0,1,...,n-1$, but excluding possibilities
given by Eq. (\ref{eq:cond1}).
The convention used here is that the quantity left of the
colon is the number of classes of the kind on the right of
the colon.

\vskip .2cm
\item[$(ii)$] $n = 3\, \mathbb{Z}$.  Eq. (\ref{eq:class3a}) now
has two solutions, thus lowering the number of possible
classes with three elements.

\begin{eqnarray}\label{eq:class5a}
2 &:& 1C_{1}^{(\nu)} ~=~
\left \{
c^{\nu}d^{2\nu}
\right \},
~~~ \nu=\mbox{$\frac{n}{3},\frac{2n}{3}$}, 
\\
n-3  &:&  3C_{1}^{(\rho)} ~=~
\left \{   
c^{\rho}d^{-\rho}, c^{-2\rho}d^{-\rho},
c^{\rho}d^{2\rho}
\right \},
~~
\rho\neq
\mbox{$\frac{n}{3},\frac{2n}{3}$},
\\
\dfrac{n^2-3n+6}{6}&:& 6C_{1}^{(\rho,\sigma)}~=~
\left \{   
c^{\rho}d^{\sigma}, c^{\sigma-\rho}d^{-\rho},
c^{-\sigma}d^{\rho-\sigma},
c^{-\sigma}d^{-\rho}, 
c^{\sigma-\rho}d^{\sigma},
c^{\rho}d^{\rho-\sigma} 
\right \},
\end{eqnarray}
where $\rho,\sigma =0,1,...,n-1$, again excluding
possibilities  given by Eq. (\ref{eq:cond1}).

\end{itemize}
\vskip .3cm

\item {{\bf{Elements with $\boldsymbol{a}$ and $\boldsymbol{a^2}$ }}}

Here we consider an element of the form $ac^{\rho}d^{\sigma}$ with
$\rho, \sigma =0,1,...,n-1$, when it is conjugated first by the
element
$g_{\alpha}=a^{\alpha}c^{\gamma}d^{\delta}$ (where $\alpha
=0,1,2$ and $\gamma, \delta=0,1,...,n-1$) and then by the
element $b$. As we had noted in the beginning of this
section conjugation by $b$ results in $a$ and $a^2$ belonging to
the same type of classes. 
When $\alpha=0$ we get

\begin{eqnarray}\label{eq:cong1b}
  g_{0} ac^{\rho}d^{\sigma}g_{0}^{-1} &=&
  ac^{\rho-\gamma-\delta} d^{\sigma + \gamma-2\delta}\\
  &=& 
  ac^{\rho+\sigma-y_{0}-3x_{0}} d^{y_{0}},
\end{eqnarray}
where we have used the redefinitions $x_{0}\equiv\delta$ and 
$y_{0}\equiv\sigma+\gamma-2\delta$, with $x_{0}, y_{0}=0,1,...,n-1$.
Applying $g_{\alpha}$ on $ac^{\rho}d^{\sigma}$ but for
$\alpha=1$ and then once again for $\alpha=2$, we find we can
summarize the three cases of $\alpha$ as 

\begin{equation}\label{eq:cong2b}
  g_{\alpha} ac^{\rho}d^{\sigma}g_{\alpha}^{-1} 
  =
  ac^{\rho+\sigma-y_{\alpha}-3x_{\alpha}} d^{y_{\alpha}},
\end{equation}
where
$y_{\alpha}=-\rho-\sigma+y_{\alpha-1}+3x_{\alpha-1}$ and
$x_{\alpha}=x_{\alpha-1}-y_{\alpha}$. Since 
both $x_{\alpha}$ and $y_{\alpha}$ take all values
$x_{\alpha},y_{\alpha}=0,1,...,n-1$, we can write the elements with a single power 
of $a$ belonging to the same class as

\begin{equation}\label{eq:cong3b}
  ac^{\tau-y-3x} d^{y}, ~~~~ x,y=0,1,...,n-1
  \ ,
\end{equation}
for a given choice of $\tau$ where $\tau \equiv \rho+\sigma$. 
Finally, the conjugation of the above
with the element $b$ will produce the elements with $a^2$, 

\begin{equation}\label{eq:cong4b}
  bac^{\tau-y-3x} d^{y}b^{-1}
  =
  a^2c^{-y}d^{y+3x-\tau}
 .
\end{equation}
So far we have new classes parametrized by $\tau$:

\begin{equation}\label{eq:cong5b}
C^{(\tau)}_{2}=
 \left \{  
  ac^{\tau-y-3x} d^{y},~
  a^2c^{-y}d^{y+3x-\tau}
  | ~~x,y=0,1,...,n-1
 \right \}, ~~  \tau=0,1,...,n-1
 \ .
\end{equation}
We can now check for conditions which would cause over counting of elements in the
same class. If we hold $y$ constant for any two
elements with the same power in $a$, we find we can over count
when $ 3 \left( x - x' \right) =0~\mathrm{mod}(n) $. We then have to 
consider once again the two separate conditions for $n$.

\begin{itemize}

\item[$(i)$]  $n \neq 3\,\mathbb{Z}$.      We do
not have to worry about redundant terms caused by $x$. Quick
inspection of  Eq.(\ref{eq:cong5b}) reveals that $\tau-y-3x$
could be replaced by redefinition $z \equiv \tau-y-3x$. 
Notice that when $y$ has been chosen, we are still free
to choose values for $x$, thus generating {\it all} possible values for
$z~\mathrm{mod}~(n)$, irrespective of initial choice of $\tau$.
Our class is therefore no longer parametrized by $\tau$:

\begin{equation}\label{eq:cong6b}
 1 ~:~
 2n^2C_{2}
 =
 \left \{  
  ac^{z} d^{y},~
  a^2c^{-y}d^{-z}
  |
 ~~  z,y=0,1,...,n-1
 \right \} 
 \ .
\end{equation}

\item[$(ii)$]  $n = 3\, \mathbb{Z}$.  Here the variable $x$, 
as mentioned, can produce over counting in the same class. So it
must be limited to $x=0,1,...,(n-3)/3$ to avoid it. Thus we allow $y$ to take on all values, 
and in doing so we need to set the values for $\tau$, which now 
parametrizes the different classes: 

\begin{equation}\label{eq:cong6bb}
 3 ~:~
 \frac{2n^2}{3}C_{2}^{(\tau)}
 =
 \{  
  ac^{\tau-y-3x} d^{y},~
  a^2c^{-y}d^{y+3x-\tau}
  |
  ~y=0,1,...,n-1, 
 ~~x=0,1,...,\mbox{$\frac{n-3}{3}$}
  \},~~~ \tau=0,1,2 \\ 
 \ .
\end{equation}

\end{itemize}
\vskip .3cm

\item {{\bf{Elements with $\boldsymbol{b}$}}}

The exercise here is similar to the one above, with the sole
exception that we begin with an element of the form
$bc^{\rho}d^{\sigma}$, but nevertheless conjugating the term
with $g_{\alpha}$ as before. Conjugations with $\alpha=0$
results in

\begin{equation}\label{eq:bconj1}
  g_{0} bc^{\rho}d^{\sigma}g_{0}^{-1} 
  =
  bc^{\rho-\gamma-\delta}d^{\sigma-\delta-\gamma}
  =
  bc^{\rho'+x}d^{x}
  \ ,
\end{equation}
where $\rho'=\rho-\sigma$, $x=\sigma-\delta-\gamma$, and
$x=0,1,...,n-1$. Now, conjugating
$bc^{\rho}d^{\sigma}$ with $g_{\alpha}$ for the cases
$\alpha=1,2$, we get one term involving $ab$ and
another $a^2b$. The final result are classes parametrized by
$\rho'$:

\begin{equation}\label{eq:cong1c}
     n ~:~
     3nC_{3}^{(\rho')}
     =
     \left \{
     bc^{\rho'+x}d^{x}, ~
     a^2bc^{-\rho'}d^{-x-\rho'},~
     abc^{-x}d^{\rho'}
     |
     ~~x=0,1,...,n-1
     \right \}, \rho'=0,1,...,n-1
     \ .
\end{equation}
The parameterization of the class comes about when one 
chooses a value for $\rho$ and
$\sigma$, the $\rho'$ value becomes fixed but not $x$. Finally
unlike previous types of classes the above is independent on 
whether or not $n=3\mathbb{Z}$.  

\end{itemize}
\vskip .3cm

This completes the derivation of the class structure 
of the group $\Delta(6n^2)$. 
\\

The results can be summarized as:
\begin{itemize}

\item[($i$)]{$n \neq  3 \,\mathbb{Z}$.}~
Five types of classes

\begin{equation}\label{eq:sum1}
1C_1, ~~~ 3C_1^{(\rho)},~~~6C_{1}^{(\rho,\sigma)}, ~~~ 2n^2C_2,
~~~ 3nC_3^{(\rho)} 
\end{equation}
adding up to $1+(n-1)+\frac{n^2-3n+2}{6}+1+n$ distinct classes.

\item[($ii$)]{$n =  3 \, \mathbb{Z}$.}~
Six types of classes

\begin{equation}\label{eq:sum2}
1C_1, ~~~ 1C_1^{(\nu)}, ~~~
3C_1^{(\rho)}, ~~~ 6C_{1}^{(\rho,\sigma)},~~~ \mbox{$\frac{2n^2\!\!}{3}$}C_2^{(\tau)},
~~~ 3nC_3^{(\rho)}, 
\end{equation}
resulting in
$1+2+(n-3)+\frac{n^2-3n+6}{6}+3+n$ different classes.

\end{itemize}

\cleqn
\section{Irreducible Representations}\label{sirreps}
We shall systematically construct all the irreducible representations of
$\Delta(6n^2)$. The explicit construction will allow us to
determine the character table by simple application of taking
the trace. Keep in mind that we shall not list the representations 
in numerical order by their dimension, but instead we shall 
list by order in which they were obtained.

\begin{itemize}
\item {{\bf{One-dimensional Representations}}}

The semidirect product produces very strong constraints on $c$
and $d$, reducing the set of possible one dimensional
representations to simply two:

\begin{eqnarray}\label{eq:onerep1}
  \mathbf{1_{1}} &:& a=b=c=d=1,\\  
  \mathbf{1_{2}} &:& a=c=d=1, ~b=-1,  
\end{eqnarray}
true regardless of $n$.

\item {{\bf{Six-dimensional Representations}}}

The method of induced representation\footnote{
Details of the method can be found described in \cite{LNR}. The coset here contains
six points:  $(e,z)$, $(a^2,z)$, $(a,z)$, $(b,z)$, $(ab,z)$,
$(a^2b,z)$. The action of $c$ and $d$ remain the same:
$cz=\eta^{l}z$, $dz=\eta^{-k-l}z$.}
shows conclusively that there exist 
six-dimensional representations. The Abelian elements ($c$ and
$d$) are $6 \times 6 $ diagonal matrices whose entries are powers ($l$ and $k$) of the $n$th root of unity, given by $\eta \equiv e^{2\pi
i/n}$ with $l,k=0,1,...,n-1$. 

\begin{equation}\label{eq:6drep}
a = \begin{pmatrix} a_{1}& 0 \\  0 & a_{2} \end{pmatrix},~~~~
b = \begin{pmatrix} 0    & {\mathbb{1}} \\  {\mathbb{1}}  & 0 \end{pmatrix},~~~~
c = \begin{pmatrix} c_{1} & 0 \\ 0 & c_{2}
\end{pmatrix},~~~~
d = \begin{pmatrix} d_{1} & 0 \\ 0 & d_{2}
\end{pmatrix}\ ,
\end{equation}
where 
\begin{eqnarray}\label{eq:6drep2}
    a_{1} = 
    \begin{pmatrix} 
    0 & 1 & 0 \\  0 & 0 & 1 \\ 1 & 0 & 0
    \end{pmatrix},~~~~
    &
    a_{2} = 
    \begin{pmatrix} 
    0 & 0 & 1 \\  1 & 0 & 0 \\ 0 & 1 & 0
    \end{pmatrix},~~~~
    \\
    c_{1} = d_{2}^{-1}=
    \begin{pmatrix} 
    \eta^{l}& 0 & 0 \\ 0 & \eta^{k} & 0 \\ 0 & 0 & \eta^{-l-k}
    \end{pmatrix},~~~~
    &
    c_{2} = d_{1}^{-1}=
    \begin{pmatrix} 
    \eta^{l+k} & 0 & 0 \\ 0 & \eta^{-l} & 0 \\ 0 & 0 & \eta^{-k}
    \end{pmatrix}~
    \ .
\end{eqnarray}
Care must be taken, in that the above terms lend themselves to
over counting. To observe this we note that there exist
similarity transformations that leave both $a$ and $b$ alone
while only exchanging the diagonal entries of both $c$ and
$d$. Two matrices that perform such transformations are 
\begin{eqnarray}\label{eq:simtra1}
V & = &
\begin{pmatrix}
a_{1} & 0 \\ 
0 & a_{1} \\ 
\end{pmatrix},~~~V^{3}=1 , \\
T & = &
\begin{pmatrix}
0 & t_{1} \\ 
t_{1} & 0 \\ 
\end{pmatrix}, ~~~T^{2}=1
\ ,
\end{eqnarray}
where
\begin{equation}
t_{1} =
\begin{pmatrix}
0 & 0 & 1\\
0 & 1 & 0\\
1 & 0 & 0\\
\end{pmatrix}
\ .
\end{equation}
When we apply the above transformation on $c$ we get the
following set of six possible labeling exchanges

\begin{equation}\label{eq:6pairs}
 \begin{pmatrix} k    \\ l    \end{pmatrix},
 \begin{pmatrix} -k-l \\ k    \end{pmatrix},
 \begin{pmatrix} l    \\ -k-l \end{pmatrix}
 \begin{pmatrix} -l   \\ -k    \end{pmatrix},
 \begin{pmatrix} k+l  \\ -l    \end{pmatrix},
 \begin{pmatrix} -k   \\ k+l \end{pmatrix}
\ .
\end{equation}
We can arrive at the above six cases by first using the
definition

\begin{equation}\label{eq:defineM}
M^{p}_{s}\equiv 
\begin{pmatrix}
-1 & -1 \\
1 & 0
\end{pmatrix}
^{p}
\begin{pmatrix}
0 & -1 \\
-1 & 0
\end{pmatrix}
^{s}
,
 ~~~~\mathrm{with} ~~~ p=0,1,2, ~~~s=0,1,
\end{equation}
so that we may write the six pairs above in the form

\begin{equation}
M^{p}_{s}
\begin{pmatrix} k \\ l \end{pmatrix}\
  \ ,
\end{equation}

for the choices of $p$ and $s$.
Notice that we can reproduce the exponents of the elements of
the classes parametrized by $(\rho,\sigma)$
(Eq. (\ref{eq:conjugate1a})) by application of
$(\rho,\sigma)M_{s}^{p}$. The six-dimensional representation
will in general be labeled by a pair of numbers $(k,l)$, but we
want to avoid the possible labeling ambiguity that arises as
mentioned above. To do so we assume the existence of a mapping 
$\widetilde{\phantom{wa}}$ that gives the standard representative:
\begin{equation}
\widetilde{\begin{pmatrix} k \\ l \end{pmatrix}} ~~ \longmapsto
~~\mathrm{either} ~~
\begin{pmatrix} k \\ l \end{pmatrix}, ~~~
\begin{pmatrix} -k-l \\ k \end{pmatrix},  ~~~
\begin{pmatrix} l \\ -k-l \end{pmatrix},~~~
\begin{pmatrix} -l \\ -k \end{pmatrix}, ~~~
\begin{pmatrix} k+l \\ -l \end{pmatrix},  ~\mathrm{or} ~~
\begin{pmatrix} -k \\ k+l \end{pmatrix}\label{mapping}
\ .
\end{equation}

We now must ask whether the six dimensional
representation is in fact irreducible or not. To begin we note
that the non-Abelian generators $a$ and $b$ both form
representation of the $S_{3}$. Taking the trace of both $a$
and $b$ we find that $Tr(a)=0$ and $Tr(b)=0$. Then
looking at the irreducible representations of $S_{3}$
(underlined), we notice
that the only possible decompositions of $a$ and $b$ is

\begin{eqnarray}\label{eq:decompa}
  a,b~ \rightarrow~ \mathbf{\underline{ 2}} +\mathbf{\underline{2}} +
      \mathbf{\underline {1}} + \mathbf{\underline{ 1'}} 
\ .
\end{eqnarray}
The above fact however does not immediately make it clear in
what way the decomposition will take place, but does show that
it is perhaps possible for some choices of $l$ and $k$. A similarity transformation may in fact 
explicitly show the above 
breakdown of $a$ and $b$ but in doing so $c$ and $d$ will in general 
be non-diagonal. In the case of non-diagonal generators $c$,
$d$ we will find conditions on the parameters $k$, $l$ that
will cause the generators $c$, $d$ to also break down according to Eq.
(\ref{eq:decompa}). It is these restrictions that limit us from
taking all possible values of $l$ and $k$ in the following list
of six-dimensional irreducible representations. The cases in
which $l=0$, $k=0$, $l+k=0~{\mathrm {mod}} (n)$, produce both three
and two-dimensional representations, while the case
$l=k=n/3,2n/3$ produces another set of two-dimensional representations.
What we will find is that there are two cases for the
six-dimensional representations:  

\begin{itemize}
\item[($i$)]{$n \neq  3 \, \mathbb{Z}$.}
The notation that follows will list the irreducible representation 
with its  
parameterization, and subsequently the conditions which would make it
reducible.

\begin{equation}
{\bf 6}_{\widetilde{(k,l)}}, ~~~~~ \mathrm{reducible~if} ~~ 
\begin{cases}
n   ~~:~~ k+ l= ~0~{\mathrm {mod}} (n)\\
n-1 ~~:~~ k = 0 ~~,~~ l \neq 0 \\
n-1 ~~:~~ l = 0 ~~,~~ k \neq 0 \\
\end{cases}
\ .
\end{equation}
Excluding the above cases for $l$ and $k$ it is easy to see
that there should be $n^2-2(n-1)-n=(n-1)(n-2)$ different
representations, however remembering Eq.~(\ref{mapping}) it is
easy to see that we need to divide this factor by six i.e.
$\frac{(n-1)(n-2)}{6}$.

\item[($ii$)]{$n =  3 \, \mathbb{Z}$.}

\begin{equation}
{\bf 6}_{\widetilde{(k,l)}}, ~~~~~ \mathrm{reducible~if} ~~ 
\begin{cases}
2~~:~~ (k,l) = (\mbox{$\frac{n}{3}$},\mbox{$\frac{n}{3}$}),
(\mbox{$\frac{2n}{3}$},\mbox{$\frac{2n}{3}$}) \\
n   ~~:~~ k+ l= ~0~{\mathrm {mod}} (n)\\
n-1 ~~:~~ k = 0 ~~,~~ l \neq 0 \\
n-1 ~~:~~ l = 0 ~~,~~ k \neq 0\\
\end{cases}
\ .
\end{equation}
Similar to the $n\neq 3\mathbb{Z}$ case with the exception of
the exclusion of the first line in the above conditions. This
means we have $n^2-2-n-2(n-1)=n(n-3)$ possibilities and again because
of the over counting issue mentioned above we really 
have $\frac{n(n-3)}{6}$ possible irreducible representations.
\end{itemize}

With Eq.~(\ref{mapping}) in mind we shall from now on omit 
the $\widetilde{\phantom{wa}}$ symbol and simply 
write ${\bf 6}_{(k,l)}$. In the case where the
restrictions are violated, it can be shown that the now reducible
representations break up into three-dimensional
and two-dimensional representations. 

\item {{\bf{Three-dimensional Representations}}}

Recall that a representation is reducible if it can be found to
be block diagonal. Looking at the six-dimensional representation, 
one cannot help but notice that the $3 \times 3$ block 
structure of all generators ($a$,$b$,$c$,$d$) is suggestive.
With this realization, we shall attempt to diagonalize $a$
and $b$ simultaneously (via similarity transformation). The 
general matrix found that diagonalizes $a$ and $b$ is of the form:

\begin{equation}\label{eq:simi1}
  S = \
\begin{pmatrix} \mathbb{1} & \mathbf{e} \\ \mathbb{1} &
\mathbf{-e} \end{pmatrix}, ~~~
\mathbf{e}=\left \{
\begin{pmatrix}
0&0&1\\
0&1&0\\
1&0&0
\end{pmatrix},
\begin{pmatrix}
0&1&0\\
1&0&0\\
0&0&1
\end{pmatrix},
\begin{pmatrix}
1&0&0\\
0&0&1\\
0&1&0
\end{pmatrix}
\right \},
\end{equation}
where $\mathbf{e}$ matrices are the order two elements 
belonging to $S_{3}$ in its three-dimensional representation. 
In block diagonalizing $a$ and $b$ we spoil the diagonal
property of $c$ and $d$. However, it may be possible to find
conditions for $k$ and $l$ that would  once again make the
representation block diagonal. In fact there exist three
possible conditions, each for the choice of $\mathbf{e}$ made.
The conditions are either
 
\begin{equation}\label{eq:condired}
k+l = 0~{\mathrm {mod}} (n),~~~~~~~~
l = 0, ~~~~~~~~
k = 0,
\end{equation}
respectively. Excluding $k=l=0$ which will be shown to give even smaller
irreducible representations, the diagonalization produces two three-dimensional
irreps listed below for the 
choice $k+l=0~{\mathrm {mod}} (n)$:
 
 \begin{equation}\label{eq:3dimrepsa}
   a_{(3_{1})}=\begin{pmatrix}0 &1 &0 \\ 0&0&1 \\
   1&0&0\end{pmatrix},~
   b_{(3_{1})}=\begin{pmatrix} 0 &0 &1 \\ 0&1&0 \\
   1&0&0\end{pmatrix},~
   c_{(3_{1})}=\begin{pmatrix} \eta^{l}&0 &0 \\ 0&\eta^{-l}&0 \\ 
   0&0&1\end{pmatrix},~
   d_{(3_{1})}=\begin{pmatrix}1 &0 &0 \\ 0&\eta^{l}&0 \\ 
   0&0&\eta^{-l}\end{pmatrix}, \\
 \end{equation}
 \begin{equation}\label{eq:3dimrepsb}
   a_{(3_{2})}~=~a_{(3_{1})},~~~~
   b_{(3_{2})}~=~-b_{(3_{1})},~~~~
   c_{(3_{2})}~=~c_{(3_{1})},~~~~
   d_{(3_{2})}~=~d_{(3_{1})}
   \ .
 \end{equation}
Once again $\eta$ is the $n$th root of unity, and
$l=1,2,...,n-1$. The other two choices in Eq.~(\ref{eq:condired}) are related to Eq.~(\ref{eq:3dimrepsa})-(\ref{eq:3dimrepsb}) by similarity transformations. So we see that 
the number of three dimensional representations in this case is 
simply $2(n-1)$.

\item {{\bf{Two-dimensional Representations}}}

The two-dimensional representations are obtained from two
distinct origins. Where one is obtained from the
conditions that make three-dimensional representation 
reducible, the other two-dimensional representations arise
directly from the six-dimensional representation. It is important to be aware 
that the first case occurs for all values of $n$.

\begin{itemize}
\item[($i$)]{$\forall n$.} In the previous
section we excluded the case when $l=k=0$, but 
if this occurs both $c$ and $d$ become nothing more than the identity matrix. 
We are then left with two three-dimensional
representations of $S_{3}$. However $S_{3}$, according to its
character table, has no irreducible three-dimensional
representations. Looking both at Eq.~(\ref{eq:decompa}) and the
character table of $S_{3}$, the only possible decompositions
are ${\mathbf {3_1}}\rightarrow{\mathbf{ \underline{2}+1_1}}$ and 
${\mathbf {3_2}}\rightarrow{\mathbf{ \underline{2}+1_2}}$. The 
difference between the $\mathbf {3_1}$ and the $\mathbf {3_2}$ 
is that the
trace of $b$ are opposite signs which simply translates in
having different one-dimensional representations in the decomposition.
In either case, they both contain the same two-dimensional
representation, ${\mathbf{ 2_1}=\underline{\mathbf 2}}$ :

\begin{equation}\label{eq:twodimrep1}
 a_{(2_{1})}= \begin{pmatrix}\om& 0 \\ 0 & 
           \om^2 \end{pmatrix},~~ 
 b_{(2_{1})}= \begin{pmatrix}0 & 1 \\ 1 & 
           0 \end{pmatrix},~~
 c_{(2_{1})}=d_{(2_{1})}= \begin{pmatrix}1 & 0 \\ 0 & 
           1 \end{pmatrix},
\end{equation}
where $\om=e^{2\pi i/3}$.
\item[($ii$)]{$n = 3 \, \mathbb{Z}$.} The three-dimensional
representation was primarily motivated by the block diagonal
structure found in the six-dimensional representation. However,
it is possible that other reductions may in fact exist.
To find these reductions we perform a similarity
transformation on both $a$ and $b$ with the following results

\begin{equation}\label{eq:simtran1}
a'=U a U^{-1}= 
  \begin{pmatrix}
  a_{(2_1)}&{0}&{0}\\
  {0}&{a_{(2_1)}}&{0}\\
  {0}&{0}&{\mathbb 1} \end{pmatrix}, 
  ~~~
  b'=UbU ^{-1} =
  \begin{pmatrix}
  b_{(2_1)}&{0}&{0}\\{0}&{b_{(2_1)} }&{0}\\
  {0}&{0}&b_{(2_1)} \end{pmatrix} 
  \ ,
\end{equation}
where all the entries are $2\times2$ matrices, the ones on the diagonal
defined in Eq.~(\ref{eq:twodimrep1}).
The last entry in $a'$ is an identity matrix to ensure that
the trace of $a'$ is identical to the trace of $a$. With the
choices specified above, one finds
constraints on $U$. When $U$, under the newly obtained
constraints, is applied to $c$ and $d$,  we find that in
general these generators are no longer diagonal. To remove off
diagonal terms, excluding the case where $k=l=0$, it's necessary that

\begin{equation}\label{eq:whatever}
k=l=n/3, 2n/3,  
\end{equation}
so that we may replace $\eta^{k}$ for $\omega$. The end
result being three two-dimensional representations, with
$k=l=n/3$ (we could have chosen $k=l=2n/3$ but it produces
equivalent results):

\begin{eqnarray}\label{eq:2dimb}
a_{(2_{2})} = a_{(2_{1})},~~
~b_{(2_2)}=b_{(2_1)},~~~c_{(2_2)}=d_{(2_2)}=~~~
\begin{pmatrix} \omega^{2}&0\\0&\omega \end{pmatrix},\\
a_{(2_{3})} = a_{(2_{1})},~~
~b_{(2_3)}=b_{(2_1)},~~~c_{(2_3)}=d_{(2_3)}=~~~
\begin{pmatrix} \omega&0\\0&\omega^{2} \end{pmatrix},\\
a_{(2_{4})} = \mathbb{1},~~
~b_{(2_4)}=b_{(2_1)},~~~c_{(2_4)}=d_{(2_4)}=
~\begin{pmatrix} \omega&0\\0&\omega^{2} \end{pmatrix}.
\end{eqnarray}
Notice that in the last representation $a_{(2_4)}$ and $b_{(2_4)}$ are not an irreducible representation
of $S_3$. Now, including the two-dimensional representation 
from the $\forall n$ case, we see
that in total we have $1+3=4$ two-dimensional
representations.
\end{itemize}

There are no more irreducible representations as can be seen by performing the following sum

\begin{equation}\label{eq:sum}
6n^2= \sum_{irrep~i} d_{i}^{2},
\end{equation} 
where $d_i$ is the dimension of the representation. Performing the sum for $n\neq3\mathbb{Z}$ we find

\begin{equation}
\frac{(n-1)(n-2)}{6}\times6^2+2(n-1)\times3^2+1\times2^2+2\times1^2=6n^2,
\end{equation}
and for $n=3\mathbb{Z}$ we have

\begin{equation}
\frac{n(n-3)}{6}\times6^2+2(n-1)\times3^2+4\times2^2+2\times1^2=6n^2.
\end{equation}

\end{itemize} 
We can derive the $\Delta(6n^2)$ character table by taking
traces over the relevant matrices, for both $n \neq 3\, \mathbb{Z}$
and $n = 3\, \mathbb{Z}$. The results are displayed in
Table~\ref{tb:characterAa}, using the classes and representations just
derived, including the restrictions on the parameters $(\rho,\sig)$
and $(k,l)$.

\begin{table}[t] 
\begin{center}
\caption{ The character tables of $\Delta{(6n^2)}$ for the
cases (a) $n \neq 3 \mathbb{Z}$ and (b) $n =3 \mathbb{Z}$. Note
$\rho$ and $\sigma$ take on different values depending on the class, 
$\tau=0,1,2$, $\nu=\frac{n}{3}$,$\frac{2n}{3}$,
$p=0,1,2$, $s=0,1$, $r=1,2$, $\om = e^{2\pi i/3}$,  and $l,k=1,2,...,n-1$. }
\label{tb:characterAa}
\begin{tabular}{cccccccc} 
\hline
\hline
\\
$\!\!$(a)$\!\!\!$& $n \neq 3 \mathbb{Z}$ & $1C_1$ & $3C_1^{(\rho)}$ &
$6C_1^{(\rho,\sigma)}$ &
$2n^2C_2$ & $3nC_3^{(\rho)}$\\
\hline
& $\mathbf{1_1}$ & $1$ & $1$ & $1$ &  $1$  & $1$  & \\ 
& $\mathbf{1_2}$ & $1$ & $1$ & $1$ &  $1$  & $-1$ & \\
& $\mathbf{2_1}$ & $2$ & $2$ & $2$ &  $-1$ & $0$  & \\
& $\mathbf{3_1}_{(l)}$ & $3$ &
$\sum_{p}\eta^{(\rho,-\rho)M^{p}_{0} 
\mbox{\tiny$\begin{pmatrix} -l\\l \end{pmatrix}$}} $   
&  $\sum_{p}\eta^{(\rho,\sigma)M^{p}_{0} 
\mbox{\tiny$\begin{pmatrix}-l\\l \end{pmatrix}$}}$ 
& $0$ & $\eta^{-\rho l}$   \\
& $\mathbf{3_2}_{(l)}$ & $3$ &  $\sum_{p}\eta^{(\rho,-\rho)M^{p}_{0} 
\mbox{\tiny$\begin{pmatrix} -l\\l \end{pmatrix}$}}$ 
& $\sum_{p}\eta^{(\rho,\sigma)M^{p}_{0} 
\mbox{\tiny$\begin{pmatrix} -l\\l \end{pmatrix}$}}$   
& $0$ & $-\eta^{-\rho l}$   \\
& $\mathbf{6}_{(k,l)}$  & $6$ &
$\sum_{p,s}\eta^{(\rho,-\rho)M^{p}_{s}
\mbox{\tiny$\begin{pmatrix} k\\l \end{pmatrix}$}}$ 
& $\sum_{p,s}\eta^{(\rho,\sigma)M^{p}_{s}
\mbox{\tiny$\begin{pmatrix} k\\l \end{pmatrix}$}}$ 
&   0  & $0$ & \\
\hline
\\
$\!\!$(b)$\!\!\!$& $n = 3 \mathbb{Z}$ & $1C_1$ &
$1C_1^{(\nu)}$&$3C_1^{(\rho)}$ & $6C_1^{(\rho,\sigma)}$ &
$\frac{2n^2}{3}C_2^{(\tau)}$ & $\!\!3nC_3^{(\rho)}\!\!$\\
\hline
& $\mathbf{1_{1}}$  & $1$ & $1$ & $1$ & $1$ & $1$ & $1$   \\ 
& $\mathbf{1_{2}}$ & $1$ & $1$ & $1$ & $1$ & $1$ & $-1$  \\
& $\mathbf{2_1}$ & $2$ & $2$ & $2$ & $2$ &  $-1$ & $0$  \\
& $\mathbf{2_2}$ & $2$ & $2$ & $2$ &
$\sum_{r}\om^{(\rho+\sigma)r}$ & 
$\sum_r \omega^{(2+\tau)r}$ & $0$  \\
& $\mathbf{2_3}$ & $2$ & $2$ & $2$ & 
$\sum_{r}\om^{(\rho+\sigma)r}$ & 
$\sum_r \omega^{(1+\tau)r}$ & $0$  \\
& $\mathbf{2_4}$ & $2$ & $2$ & $2$ & 
$\sum_{r}\om^{(\rho+\sigma)r}$  & 
$\sum_r \omega^{(\tau) r}$ & $0$  \\
& $\mathbf{3_1}_{(l)}$ & $3$ & $\sum_{p}\eta^{(\nu,-\nu)M^{p}_{0} 
\mbox{ \tiny $\begin{pmatrix} -l \\l \end{pmatrix}$ }}$  
& $\sum_{p}\eta^{(\rho,-\rho)M^{p}_{0} 
\mbox{\tiny $\begin{pmatrix} -l \\l \end{pmatrix}$}} $   
&  $\sum_{p}\eta^{(\rho,\sigma)M^{p}_{0} 
\mbox{\tiny $\begin{pmatrix} -l \\l \end{pmatrix}$}}$
& $0$ & $\!\!\eta^{-\rho l}\!\!$   \\
& $\mathbf{3_2}_{(l)}$ & $3$ &  $\sum_{p}\eta^{(\nu,-\nu)M^{p}_{0} 
\mbox{\tiny $\begin{pmatrix} -l \\l \end{pmatrix}$}}$
&  $\sum_{p}\eta^{(\rho,-\rho)M^{p}_{0} 
\mbox{\tiny $\begin{pmatrix} -l \\l \end{pmatrix}$}}
$ & $\sum_{p}\eta^{(\rho,\sigma)M^{p}_{0} 
\mbox{\tiny $\begin{pmatrix} -l \\l \end{pmatrix}$}}$
& $0$ & $\!\!-\eta^{-\rho l}\!\!$   \\
& $\mathbf{6}_{(k,l)}$  & $6$ &
$\sum_{p,s}\eta^{(\nu,-\nu)M^{p}_{s}
\mbox{\tiny $\begin{pmatrix} k\\l \end{pmatrix}$}}
$&$\sum_{p,s}\eta^{(\rho,-\rho)M^{p}_{s}
\mbox {\tiny $\begin{pmatrix} k\\l \end{pmatrix}$}} $ &
$\sum_{p,s}\eta^{(\rho,\sigma)M^{p}_{s}
\mbox{ \tiny $\begin{pmatrix} k\\l \end{pmatrix}$}}$ 
&   0  & $0$ \\
\hline
\hline
\end{tabular}
\end{center}
\end{table}

Looking at the character table above, notice that the
entries for the class $C_{1}$, $C_{1}^{(\rho)}$, and
$C_{1}^{(\nu)}$ can all be obtained by suitable
substitutions in the parameters of $C^{(\rho,\sigma)}_{1}$. A
fact that will be a source of great simplification in the next
section. 

\cleqn
\section{Kronecker Products}\label{product}

In order to build models, it is important to know which
products of representations will lead to invariant quantities,
associated with the singlet representations.
So it is necessary to determine all the Kronecker products of the
irreducible representations. In general the  Kronecker products
are obtained by

\beqn \bf{r} \otimes \bf{s} & = & \sum_{\bf t}d(\bf
{r},\bf{s},\bf{t}) ~ \bf{t}\ , \eeqn 
where $\bf r$, $\bf s$, and $\bf t$ are irreducible
representations, and the sum is taken over all 
irreducible representations. One approach of solving for the the integer numerical factors $d(\bf
{r},\bf{s},\bf{t})$ would be to calculate it from the character
table by using 

\beqn
d(\bf {r},\bf{s},\bf{t}) & = & \frac{1}{N} \sum_{i} n_i \cdot \chi^{[\bf r]}_i \, \chi^{[\bf s]}_i
\, \ol{\chi}^{[\bf t]}_i\ , \label{decomp} \eeqn
where $N$ is the order of the group; $i$ labels a class of $n_i$ elements and
character $\chi^{}_{i}$.   $\ol{\chi}_i$~denotes the complex conjugate
character. Another approach, discussed in Appendix
\ref{calcdetail}, obtains the same results, by
explicitly building the products. Below we shall only summarize the results for all Kronecker products,
relegating the details to Appendix B.

\begin{itemize}

\item[$(i)$] {$n \neq  3 \, \mathbb{Z}$.}

Looking at Table \ref{tb:characterAa}a, we 
note that there are: $2$ one-dimensional, $1$ 
two-dimensional, $2(n-1)$ three-dimensional, and
$\frac{(n-1)(n-2)}{6}$ six-dimensional representations.

\begin{eqnarray}
 {\bf  1_2} \otimes {\bf 1_{2}} &=& {\bf 1_{1}} \notag \\
 {\bf  1_2} \otimes {\bf 2_1} &=& {\bf 2_1} \notag \\
 {\bf  1_2} \otimes {\bf 3_{1}}_{(l)} &=& {\bf 3_{2}}_{(l)} \notag \\
 {\bf  1_2} \otimes {\bf 3_{2}}_{(l)} &=& {\bf 3_{1}}_{(l)} \notag \\
 {\bf  1_2} \otimes {\bf 6}_{(k,l)} &=& {\bf 6}_{(k,l)} \notag\\
 {\bf  2_1} \otimes {\bf 2_1} &=& {\bf 1_{1}}+{\bf 1_{2}}+{\bf
 2_1} \notag \\
 {\bf  2_1} \otimes {\bf 3_{1}}_{(l)} &=& {\bf 3_{1}}_{(l)}+{\bf 3_{2}}_{(l)} \notag \\
 {\bf  2_1} \otimes {\bf 3_{2}}_{(l)} &=& {\bf 3_{1}}_{(l)}+{\bf 3_{2}}_{(l)} \notag \\
 {\bf  2_1} \otimes {\bf 6}_{(k,l)} &=& 
 {\bf 6}_{(k,l)}+ {\bf 6}_{(k,l)}
 \notag \\
 {\bf  3_1}_{(l)} \otimes {\bf 3_1}_{(l')} &=&
 {\bf 3_1}_{(l+l')}+{\bf 6}_{\widetilde{(l,-l')}} \notag \\
 {\bf  3_1}_{(l)} \otimes {\bf 3_2}_{(l')} &=&
 {\bf 3_2}_{(l+l')}+{\bf 6}_{\widetilde{(l,-l')}} \notag \\
 {\bf  3_1}_{(l)} \otimes {\bf 6}_{(k',l')} &=&
 {\bf 6}_{{\widetilde{\mbox{\tiny{$\begin{pmatrix}k' \\
 l'-l\end{pmatrix}$}}}}}
 +
 {\bf 6}_{{\widetilde{\mbox{\tiny{$\begin{pmatrix} k'-l \\
 l'+l\end{pmatrix}$}}}}}
 +
 {\bf 6}_{{\widetilde{\mbox{\tiny{$\begin{pmatrix} l+k' \\
 l'\end{pmatrix}$}}}}} \notag \\
 {\bf  3_2}_{(l)} \otimes {\bf 3_2}_{(l')} &=&
 {\bf 3_1}_{(l+l')}+{\bf 6}_{\widetilde{(l,-l')}} \notag \\
 {\bf  3_2}_{(l)} \otimes {\bf 6}_{(k',l')} &=&
 {\bf 6}_{{\widetilde{\mbox{\tiny{$\begin{pmatrix}k' \\
 l'-l\end{pmatrix}$}}}}}
 +
 {\bf 6}_{{\widetilde{\mbox{\tiny{$\begin{pmatrix} k'-l \\
 l'+l\end{pmatrix}$}}}}}
 +
 {\bf 6}_{{\widetilde{\mbox{\tiny{$\begin{pmatrix} l+k' \\
 l'\end{pmatrix}$}}}}} \notag \\
 {\bf  6}_{(k,l)} \otimes {\bf 6}_{(k',l')} &=&
 \sum_{p,s} {\bf 6}_{\left ({\widetilde{\mbox{\tiny{$\begin{pmatrix} k \\ l \end{pmatrix}
 +
 M^p_s\begin{pmatrix} k' \\ l' \end{pmatrix}$}}}}
 \right )
 } \notag
\end{eqnarray}

\vspace{0mm}

\item[$(ii)$] {$n =  3 \, \mathbb{Z}$.}

Now in Table
\ref{tb:characterAa}b, we note that
there are: $2$ one-dimensional, $4$ 
two-dimensional, $2(n-1)$ three-dimensional, and
$\frac{n(n-1)}{6}$ six-dimensional representations.

\vspace{0mm}

\begin{eqnarray}
 {\bf  1_2} \otimes {\bf 1_{2}} &=& {\bf 1_{1}} \notag \\
 {\bf  1_2} \otimes {\bf 2_1} &=& {\bf 2_1} \notag \\
 {\bf  1_2} \otimes {\bf 2_2} &=& {\bf 2_2} \notag \\
 {\bf  1_2} \otimes {\bf 2_3} &=& {\bf 2_3} \notag \\
 {\bf  1_2} \otimes {\bf 2_4} &=& {\bf 2_4} \notag \\
 {\bf  1_2} \otimes {\bf 3_{1}}_{(l)} &=& {\bf 3_{2}}_{(l)} \notag \\
 {\bf  1_2} \otimes {\bf 3_{2}}_{(l)} &=& {\bf 3_{1}}_{(l)} \notag \\
 {\bf  1_2} \otimes {\bf 6}_{(k,l)} &=& {\bf 6}_{(k,l)} \notag \\
 {\bf  2_1} \otimes {\bf 2_1} &=& {\bf 1_{1}}+{\bf 1_{2}}+{\bf 2_1} \notag \\ 
 {\bf  2_1} \otimes {\bf 2_2} &=& {\bf 2_3}+{\bf 2_4} \notag \\ 
 {\bf  2_1} \otimes {\bf 2_3} &=& {\bf 2_2}+{\bf 2_4} \notag \\ 
 {\bf  2_1} \otimes {\bf 2_4} &=& {\bf 2_2}+{\bf 2_3} \notag \\ 
 {\bf  2_1} \otimes {\bf 3_{1}}_{(l)} &=& {\bf 3_{1}}_{(l)}+{\bf 3_{2}}_{(l)} \notag \\
 {\bf  2_1} \otimes {\bf 3_{2}}_{(l)} &=& {\bf 3_{1}}_{(l)}+{\bf 3_{2}}_{(l)} \notag \\
 {\bf  2_1} \otimes {\bf 6}_{(k,l)} &=& 
 {\bf 6}_{(k,l)}+ {\bf 6}_{(k,l)} \notag  \end{eqnarray} \begin{eqnarray}
 {\bf  2_2} \otimes {\bf 2_2} &=& {\bf 1_{1}}+{\bf 1_{2}}+{\bf 2_2} \notag \\ 
 {\bf  2_2} \otimes {\bf 2_3} &=& {\bf 2_1}+{\bf 2_4} \notag \\ 
 {\bf  2_2} \otimes {\bf 2_4} &=& {\bf 2_1}+{\bf 2_3} \notag \\ 
 {\bf  2_2} \otimes {\bf 3_{1}}_{(l)} &=& {\bf
 6}_{
	   {\widetilde {\mbox
           {\tiny 
	   $\begin{pmatrix}
           2n/3-l \\ 2n/3+l
           \end{pmatrix}
	   $
	   }}}}
 \notag \\
 {\bf  2_2} \otimes {\bf 3_{2}}_{(l)} &=& {\bf
 6}_{
	   {\widetilde {\mbox
           {\tiny 
	   $\begin{pmatrix}
           2n/3-l \\ 2n/3+l
           \end{pmatrix}
	   $
	   }}}}
 \notag \\
 {\bf  2_2} \otimes {\bf 6}_{(k,l)} &=& 
          {\bf 6}_{
	   {\widetilde {\mbox
           {\tiny 
	   $\begin{pmatrix}
	   k+n/3 \\l+n/3
           \end{pmatrix}
	   $
	   }}}}
	  + 
          {\bf 6}_{
	   {\widetilde {\mbox
           {\tiny 
	   $\begin{pmatrix}
	   k+2n/3 \\ l+2n/3 
           \end{pmatrix}
	   $
	   }}}}
	  \notag \\
 {\bf  2_3} \otimes {\bf 2_3} &=& 
          {\bf 1_{1}}+{\bf 1_{2}}+{\bf 2_3} \notag \\ 
 {\bf  2_3} \otimes {\bf 2_4} &=& {\bf 2_1}+{\bf 2_2} \notag \\ 
 {\bf  2_3} \otimes {\bf 3_{1}}_{(l)} &=& 
          {\bf 6}_{
	   {\widetilde {\mbox
           {\tiny 
	   $\begin{pmatrix}
	   {n/3}-l\\{n/3}+l
           \end{pmatrix}
	   $
	   }}}}
	  \notag \\
 {\bf  2_3} \otimes {\bf 3_{2}}_{(l)} &=& 
          {\bf 6}_{
	   {\widetilde {\mbox
           {\tiny 
	   $\begin{pmatrix}
	  {n/3}-l\\{n/3}+l
           \end{pmatrix}
	   $
	   }}}}
	  \notag \\
 {\bf  2_3} \otimes {\bf 6}_{(k,l)} &=& 
          {\bf 6}_{
	   {\widetilde {\mbox
           {\tiny 
	   $\begin{pmatrix}
	   k+{n}{3}\\ l+{n/3}
           \end{pmatrix}
	   $
	   }}}}
	  + 
          {\bf 6}_{
	   {\widetilde {\mbox
           {\tiny 
	   $\begin{pmatrix}
	   k+2n/3\\l+{2n/3}
           \end{pmatrix}
	   $
	   }}}}
	  \notag \\
 {\bf  2_4} \otimes {\bf 2_4} &=& 
          {\bf 1_{1}}+{\bf 1_{2}}+{\bf 2_4} \notag \\ 
 {\bf  2_4} \otimes {\bf 3_{1}}_{(l)} &=& 
          {\bf 6}_{
	   {\widetilde {\mbox
           {\tiny 
	   $\begin{pmatrix}
	   {n/3}-l\\{n/3}+l
           \end{pmatrix}
	   $
	   }}}}
	  \notag \\
 {\bf  2_4} \otimes {\bf 3_{2}}_{(l)} &=& 
          {\bf 6}_{
	   {\widetilde {\mbox
           {\tiny 
	   $\begin{pmatrix}
	   {n/3}-l\\ {n/3}+l
           \end{pmatrix}
	   $
	   }}}}
	  \notag \\
 {\bf  2_4} \otimes {\bf 6}_{(k,l)} &=& 
          {\bf 6}_{
	   {\widetilde {\mbox
           {\tiny 
	   $\begin{pmatrix}
	   k+{n/3}\\l+{n/3}
           \end{pmatrix}
	   $
	   }}}}
	  + 
          {\bf 6}_{
	   {\widetilde {\mbox
           {\tiny 
	   $\begin{pmatrix}
	   k+{2n/3}\\l+{2n/3}
           \end{pmatrix}
	   $
	   }}}}
	  \notag \\
 {\bf  3_1}_{(l)} \otimes {\bf 3_1}_{(l')} &=&
 {\bf 3_1}_{(l+l')}+{\bf 6}_{\widetilde{(l,-l')}} \notag \\
 {\bf  3_1}_{(l)} \otimes {\bf 3_2}_{(l')} &=&
 {\bf 3_2}_{(l+l')}+{\bf 6}_{\widetilde{(l,-l')}} \notag \\
 {\bf  3_1}_{(l)} \otimes {\bf 6}_{{(k',l')}} &=&
 {\bf 6}_{{\widetilde{\mbox{\tiny{$\begin{pmatrix}k' \\
 l'-l\end{pmatrix}$}}}}}
 +
 {\bf 6}_{{\widetilde{\mbox{\tiny{$\begin{pmatrix} k'-l \\
 l'+l\end{pmatrix}$}}}}}
 +
 {\bf 6}_{{\widetilde{\mbox{\tiny{$\begin{pmatrix} l+k'\\
 l'\end{pmatrix}$}}}}} \notag \\
 {\bf  3_2}_{(l)} \otimes {\bf 3_2}_{(l')} &=&
 {\bf 3_1}_{(l+l')}+{\bf 6}_{\widetilde{(l,-l')}} \notag \\
 {\bf  3_2}_{(l)} \otimes {\bf 6}_{(k',l')} &=&
 {\bf 6}_{{\widetilde{\mbox{\tiny{$\begin{pmatrix}k' \\
 l'-l\end{pmatrix}$}}}}}
 +
 {\bf 6}_{{\widetilde{\mbox{\tiny{$\begin{pmatrix} k'-l \\
 l'+l\end{pmatrix}$}}}}}
 +
 {\bf 6}_{{\widetilde{\mbox{\tiny{$\begin{pmatrix} l+k'\\
 l'\end{pmatrix}$}}}}} \notag \\
 {\bf  6}_{(k,l)} \otimes {\bf 6}_{(k',l')} &=&
 \sum_{p,s} {\bf 6}_{\left ({\widetilde{\mbox{\tiny{$\begin{pmatrix} k \\ l \end{pmatrix}
 +
 M^p_s\begin{pmatrix} k' \\ l' \end{pmatrix}$}}}}
 \right )
 } \notag
\end{eqnarray}

\end{itemize}

As a final note, occasionally the right hand side of a Kronecker product will
contain terms that are reducible. As a general rule, one can go through
Section~\ref{sirreps}, in particular Eqs.~(\ref{eq:condired}) and
(\ref{eq:whatever}), and determine  
how certain representations become reducible and what irreducible
representations they break up into. Below we list all possible cases
explicitly:

\begin{equation}
{\bf 6}_{(-l,l)} , ~{\bf 6}_{(0,-l)} ,~ {\bf 6}_{(l,0)} ~ \rightarrow ~ 
{\bf 3_{1} }_{(l)} + {\bf 3_{2} }_{(l)}, 
\end{equation}

\begin{equation}
{\bf 3_{r} }_{(l)} ~\rightarrow~ {\bf 2_1} + {\bf 1}_r,~~~~~~r=1,2,
\end{equation}

and additionally for the case $n = 3{\mathbb Z}$ 
\begin{equation}
{\bf 6}_{(n/3 , n/3)} ,~ {\bf 6}_{(2n/3 , 2n/3)} ~\rightarrow~ {\bf 
2_2} +{\bf 2_3} +{\bf 2_4}. 
\end{equation}

\cleqn
\section{Conclusion}\label{outlook}

In order to carry out a systematic investigation of how a
non-Abelian finite group can be used to successfully explain the remarkable
tri-bimaximal structure in the neutrino mixing, a thorough knowledge of the
finite group is essential. 
In this article we have studied the class structure of the group
$\Delta(6n^2)$ and its irreducible representations. Introducing a compact
notation which is valid for arbitrary $n$, the Kronecker products as
well as the Clebsch-Gordan coefficients have been derived. Our results can be
used to easily calculate the group invariants (e.g. trilinear Yukawa
couplings) of one particular group defined by the value for $n$, in particular
going beyond the smallest groups with $n=1,2$. On the other
hand, comparing cases with different $n$ might unveil important common
or distinguishing features which are crucial in building viable flavor
models. Our study complements similar work~\cite{LNR} on the
group~$\Delta(3n^2)$.

\section*{Acknowledgments}

The work of CL is supported by the University of Florida 
through the Institute for Fundamental Theory and that of JAE is
supported by the Department of Energy Grant No.
DE-FG02-97ER41029. We are grateful for 
helpful discussions with Salah Nasri and Pierre Ramond.

\section*{Appendix}

\appendix

\section{Labeling the Six-Dimensional Representations}
\label{standard}

It was shown in Ref.~\cite{LNR} that the labeling of the three-dimensional
irreducible representations of $\Delta(3n^2)$ by $(k,l)$ is
ambiguous. A possible resolution by choosing a particular standard
representative was presented in three tables. 
A similar ambiguity arises in the case of $\Delta(6n^2)$ where the
six-dimensional irreducible representations are labeled by $(k,l)$. Due to
Eq.~(\ref{mapping}) we obtain the following sets of possible standard
representatives, indicated by dots on the $n \times n$ grid. We distinguish between three cases: $(i_1)$ $n=3z + 1$ with $z\in \mathbb{Z}$,
$(i_2)$ $n=3z + 2$, and $(ii)$ $n=3z$.

\begin{itemize}
\item[($i_1$)] $n = 3 z +1$.\vspace{-4mm}
\begin{center}
{\small \begin{tabular}{c||c|cccccccccccc}
$(k,l)$ & 0 &1& $\cdot$ &$\cdot$ & $z$ &$\cdot$ &$\cdot$ &$\cdot$ &$2z$
&$\cdot$ & $\cdot$ &$\cdot$  & $3z$ \\ \hline\hline
0& &  &  & &  &  & &
& &  & &   & \\ \hline
1& & $\cdot$ & $\cdot$ &$\cdot$ &  $\cdot$ &  $\cdot$ &$\cdot$ & $\cdot$ & &   & &  &\\
$\cdot$ & & $\cdot$ & $\cdot$ &$\cdot$ &  $\cdot$ &  $\cdot$ & $\cdot$ &
& &   & &  &\\
$\cdot$ & & $\cdot$ & $\cdot$ &$\cdot$ &  $\cdot$ & $\cdot$ & &
& &   & &  &\\
$z$& &$\cdot$ &  $\cdot$ & $\cdot$ & $\cdot$ & & & &   & &  &&\\
$\cdot$ &&&&&&&&&&&&& \\
$\cdot$ &&&&&&&&&&&&& \\
$\cdot$ &&&&&&&&&&&&& \\
$2z$ & &&&&&&&&& & & &  \\
$\cdot$&&&&&&&&&&  &  &  &  \\
$\cdot$&&&&&&&&&&  & & &  \\
$\cdot$&&&&&&&&&& & & &  \\
$3z$&&&&&&&&&&&&&\\
\end{tabular}}
\end{center}

\vspace{1mm}

\item[($i_2$)] $n = 3 z +2$.\vspace{-4mm}
\begin{center}
{\small \begin{tabular}{c||c|ccccccccccccl}
$(k,l)$ & 0 &1& $\cdot$ &$\cdot$ & $z$ &$\cdot$ &$\cdot$ &$\cdot$ &$2z$
&$\cdot$ & $\cdot$ &$\cdot$  & $3z$ & $\!\!\!\!3z\!\!+\!\!1$ \\ \hline\hline
0& &  &  & &  &   & &
& &   & &   & & \\ \hline
1& & $\cdot$ & $\cdot$ &$\cdot$ &  $\cdot$ &  $\cdot$ &$\cdot$ & $\cdot$&
$\cdot$  & &   & &  &\\
$\cdot$ & & $\cdot$ & $\cdot$ &$\cdot$ &  $\cdot$ &  $\cdot$ & $\cdot$
&$\cdot$ & & &   & &  &\\
$\cdot$ & & $\cdot$ & $\cdot$ &$\cdot$ &  $\cdot$ & $\cdot$ &$\cdot$ & &
& &   & &  &\\
$z$& &$\cdot$ &  $\cdot$ & $\cdot$ & $\cdot$ &$\cdot$ & & & &   & &  &&\\
$\cdot$ &&&&&&&&&&&&& \\
$\cdot$ &&&&&&&&&&&&& \\
$\cdot$ &&&&&&&&&&&&& \\
$2z$ & &&&&&&&&& & & &  \\
$\cdot$ & &&&&&&&&& & & &  \\
$\cdot$&&&&&&&&&& &  &  &  &  \\
$\cdot$&&&&&&&&& &&  &  &  &  \\
$3z$&&&&&&&& &&&  &  &  &  \\
$3z+1$&&&&&&&&&&&&&&\\
\end{tabular}}
\end{center}

\item[($ii$)] $n = 3 z $.\vspace{-4mm}
\begin{center}
{\small \begin{tabular}{c||c|ccccccccccl}
$(k,l)$ & 0 &1& $\cdot$ &$\cdot$ & $z$ &$\cdot$ &$\cdot$ &$\cdot$ &$2z$
&$\cdot$ & $\;\,\cdot\;\,$ & $\!\!\!\!3z\!\!-\!\!1$ \\ \hline\hline
0& &   & &  &   & &
& &   & &   & \\ \hline
1& &  $\cdot$ &$\cdot$ &  $\cdot$ &  $\cdot$ &$\cdot$ & $\cdot$ & &   & &  &\\
$\cdot$ & &  $\cdot$ &$\cdot$ &  $\cdot$ &  $\cdot$ & $\cdot$ &
& &   & &  &\\
$\cdot$ & & $\cdot$ & $\cdot$ &  $\cdot$ & $\cdot$ & &
& &   & &  &\\
$z$& &$\cdot$ &   $\cdot$ & $\cdot$ & & & &   & &  &&\\
$\cdot$ &&&&&&&&&&&& \\
$\cdot$ &&&&&&&&&&&& \\
$\cdot$ &&&&&&&&&&&& \\
$2z$ & &&&&&&&& &  &  &  \\
$\cdot$&&&&&&&&&&  &  &  \\
$\cdot$&&&&&&&&&&  &  &  \\
$3z-1$&&&&&&&&&&& &\\
\end{tabular}}
\end{center}

\end{itemize}

\cleqn
\section{Product of Irreducible Representations: Details}\label{calcdetail}

There are two ways of obtaining the quantities
$d({\bf r},{\bf s},{\bf t})$. The first method involves the use
of Eq.~(\ref{decomp}) in conjunction with the Character 
Table~\ref{tb:characterAa}. The calculations will not be
explicitly performed as in \cite{LNR}. The second method relies
on deriving the actual representations of the products explicitly, 
providing us immediately the Kronecker products and also the 
Clebsch-Gordan coefficients.

\begin{itemize}

\item[$(B.i)$]  {Products via the Character Table.}

First note that the classes
$C_1^{(\rho)}$, and $C_1$, (also $C_1^{(\nu)}$ for the $n=3\mathbb Z$
case) can all be obtained from $C_1^{(\rho,\sigma)}$, by the appropriate
choices of the parameters $\rho$ and $\sigma$. For the case that
$n \neq 3\mathbb Z$ we can formally express the sum over classes
$C_1^{(\rho,\sigma)}$ as 

\begin{equation}
\sum_{C_1^{(\rho,\sig)}}  =  \frac{1}{6} \left(
\sum_{\rho,\sig=0}^{n-1}~~~-~~~
\sum_{\substack{\rho+\sig~=~0~\mathrm{mod}(n)\\2\rho-\sigma~=~0~\mathrm{mod}(n)\\
\rho-2\sigma~=~0~\mathrm{mod}(n)}}
  ~~~-~~~  \sum_{\rho=\sig=0} \right).\label{c1sum}
\end{equation}
The second term on the right hand side 
contains the three conditions of Eq.~(\ref{eq:cond1}) on $\rho$ and $\sigma$, all of which lead to the
same class $C_1^{(\rho)}$ of three elements.
The factor of $1/6$ comes from the six distinct elements within one class, see Eq.~(\ref{eq:class1a}). Looking at
Eq.~(\ref{decomp}) we find that we may now write the terms
involving the classes $C_1$, $C_1^{(\rho)}$, and $C_1^{(\rho,\sigma)}$ in a
compact way  

\begin{equation}\chi^{[{\bf r}]}_{C_1} \,
\chi^{[{\bf s}]}_{C_1} \, \ol{\chi}^{[{\bf t}]}_{C_1}  ~+
\sum_{C_1^{(\rho)}} 3 \cdot \chi^{[{\bf r}]}_{C_1^{(\rho)}}
\, \chi^{[{\bf s}]}_{C_1^{(\rho)}} 
\ol{\chi}^{[{\bf t}]}_{C_1^{(\rho)}} ~+
\sum_{C_1^{(\rho,\sig)}} 6 \cdot \chi^{[{\bf r}]}_{C_1^{(\rho,\sig)}}
\, \chi^{[{\bf s}]}_{C_1^{(\rho,\sig)}} \,
\ol{\chi}^{[{\bf t}]}_{C_1^{(\rho,\sig)}} =
\sum_{\rho,\sig=0}^{n-1}   \chi^{[{\bf r}]}_{C_1^{(\rho,\sig)}} \,
\chi^{[{\bf s}]}_{C_1^{(\rho,\sig)}} \,
\ol{\chi}^{[{\bf t}]}_{C_1^{(\rho,\sig)}}.~~~
\end{equation}

The above line means that the classes $C_1$, $C_1^{(\rho)}$ and $C_1^{(\rho,\sig)}$
can be neatly written as a sum over all $n^2$ pairs 
of $(\rho,\sig)$. For the case that $n=3\mathbb Z$ a similar
expression occurs
\begin{eqnarray}
&&\hspace{-10mm}\chi^{[{\bf r}]}_{C_1} \,
\chi^{[{\bf s}]}_{C_1} \, \ol{\chi}^{[{\bf t}]}_{C_1}  ~+
\sum_{C_1^{(\nu)}} \chi^{[{\bf r}]}_{C_1^{(\nu)}}
\, \chi^{[{\bf s}]}_{C_1^{(\nu)}} 
\ol{\chi}^{[{\bf t}]}_{C_1^{(\nu)}} ~+
\sum_{C_1^{(\rho)}} 3 \cdot \chi^{[{\bf r}]}_{C_1^{(\rho)}}
\, \chi^{[{\bf s}]}_{C_1^{(\rho)}} 
\ol{\chi}^{[{\bf t}]}_{C_1^{(\rho)}} ~+
\sum_{C_1^{(\rho,\sig)}} 6 \cdot \chi^{[{\bf r}]}_{C_1^{(\rho,\sig)}}
\, \chi^{[{\bf s}]}_{C_1^{(\rho,\sig)}} \,
\ol{\chi}^{[{\bf t}]}_{C_1^{(\rho,\sig)}} \notag \\ &&\hspace{80mm}=~
\sum_{\rho,\sig=0}^{n-1}   \chi^{[{\bf r}]}_{C_1^{(\rho,\sig)}} \,
\chi^{[{\bf s}]}_{C_1^{(\rho,\sig)}} \,
\ol{\chi}^{[{\bf t}]}_{C_1^{(\rho,\sig)}}.
\end{eqnarray}
Applying these relations, it is straightforward, though still
tedious, to determine the integers $d({\bf r},{\bf s},{\bf t})$ of
Eq.~(\ref{decomp}).

\item[$(B.ii)$]  {Explicitly Building Products of Irreducible Representations.}

Here we form explicitly the representations produced by taking
products of irreducible representations. The investigation will 
be performed in some detail on the two separate cases for $n$.

     \begin{itemize}
     \item[$(i)$] {$n\neq3\mathbb Z$.}
     We first need to form several vector spaces, one for each
     irreducible representation (with the exception of the
     trivial one-dimensional representations). The spaces
     become useful when we explicitly show how each transforms
     according to the generators $a$, $b$, and $c$:\footnote{We omit the
     discussion of the generator $d$ as it is not independent 
     of $a$, $b$, and $c$, see Eq.~(\ref{eq:pres2})}

     \begin{eqnarray} \label{eq:repsa}
     {\mathbf 6}_{(k,l)} &:&~~~~
     \begin{pmatrix} x_1 \\ x_2 \\ x_3 \\ x_4 \\ x_5 \\ x_6
     \end{pmatrix}
     \mapsto
     \begin{pmatrix} x_2 \\ x_3 \\ x_1 \\ x_6 \\ x_4 \\ x_5
     \end{pmatrix}_{a}
     ,~~~
     \begin{pmatrix} x_4 \\ x_5 \\ x_6 \\ x_1 \\ x_2 \\ x_3
     \end{pmatrix}_{b}
     ,~~~
     \begin{pmatrix} \eta^{l}x_1 \\ \eta^{k}x_2 \\ \eta^{-k-l}x_3
     \\ \eta^{k+l}x_4 \\ \eta^{-l}x_5 \\ \eta^{-k}x_6
     \end{pmatrix}_{c}, \notag 
     \\
     {\mathbf 3_1}_{(l)} &:&~~~~
     \begin{pmatrix} x_1 \\ x_2 \\ x_3 
     \end{pmatrix}
     \mapsto
     \begin{pmatrix} x_2 \\ x_3 \\ x_1 
     \end{pmatrix}_{a}
     ,~~~
     \begin{pmatrix} x_3 \\ x_2 \\ x_1 
     \end{pmatrix}_{b}
     ,~~~
     \begin{pmatrix} \eta^{l}x_1 \\ \eta^{-l}x_2 \\ x_3
     \end{pmatrix}_{c},
     \\
     {\mathbf 3_2}_{(l)} &:&~~~~
     \begin{pmatrix} x_1 \\ x_2 \\ x_3 
     \end{pmatrix}
     \mapsto
     \begin{pmatrix} x_2 \\ x_3 \\ x_1 
     \end{pmatrix}_{a}
     ,~~~
     \begin{pmatrix} -x_3 \\ -x_2 \\ -x_1 
     \end{pmatrix}_{b}
     ,~~~
     \begin{pmatrix} \eta^{l}x_1 \\ \eta^{-l}x_2 \\ x_3
     \end{pmatrix}_{c}, \notag
     \\
     {\mathbf 2_1} &:&~~~~
     \begin{pmatrix} x_1 \\ x_2 
     \end{pmatrix}
     \mapsto
     \begin{pmatrix} \omega x_1 \\ \omega^2 x_2 
     \end{pmatrix}_{a}
     ,~~~
     \begin{pmatrix} x_2 \\ x_1 
     \end{pmatrix}_{b}
     ,~~~
     \begin{pmatrix} x_1 \\ x_2 
     \end{pmatrix}_{c}.\notag
     \end{eqnarray} 
     The first product below will demonstrate the procedure in
     defining the product spaces, but its worth noting that we
     always begin with one term. The importance of this is that 
     each term will require us to refer to the above
     list to determine how that first term transforms under the
     action of any generator. These transformations will in
     general lead to new terms that should also be included in
     a product space. In this way we build a set of terms
     that together define a space of an irreducible
     representation.

         \begin{itemize}
          \item[$\bullet$]{${\mathbf2_1} \otimes {\mathbf 2_1}={\mathbf
	  1_1}+ {\mathbf 1_2} +{\mathbf 2_1}$.}
	 We demonstrate here how to construct the
	 representations of the products, keeping in mind that the 
         remainder products will follow along similar lines. 
         First we start with a 
	 vector $x$, which transforms according to 
	 the first ${\mathbf 2_1}$, and then we include $y$, which
	 transforms as according to the other $\mathbf 2_1$.
         Picking any term that is a product of these two spaces, 
         we prefer to start with $x_1 y_1$, we note by
         Eq.~(\ref{eq:repsa}) that 

	 \begin{equation}
	 x_1 y_1 \mapsto_a w^2 x_1 y_1,~~\mapsto_b
	 x_2 y_2,~~\mapsto_c x_1 y_1 .
	 \end{equation}
	 The transformations have led us to conclude that
         whatever vector space $x_1 y_1$ may occupy, $x_2 y_2$ must
         also. Thus similarly $x_2 y_2$ maps according to

	 \begin{equation}
	 x_2 y_2 \mapsto_a w x_2 y_2,~~\mapsto_b
	 x_1 y_1,~~\mapsto_c x_2 y_2 .
	 \end{equation}
         We now see that no new term has been produced, we can
         safely conclude that we have all the terms of this new
         vector space, and so we once again look at Eq.~(\ref{eq:repsa})
         and find that

	 \begin{equation}
	 {\mathbf 2_1} ~~:~~
	 \begin{pmatrix}
	 x_2 y_2 \\ x_1 y_1
	 \end{pmatrix}, 
	 \end{equation}
         as can be verified by applying the actions of the
         generators. Similarly we may look at the remainder possible
	 combinations $x_1 y_2$ and $x_2 y_1$. If we
	 perform the same analysis, we may guess that we 
         have another two-dimensional representation. However 
         on closer inspection we would find that this
         two-dimensional representation is actually reducible. 
	 The two terms can be shown to occur in two linear combinations 
         that form vector spaces of one-dimensional irreducible 
         representations: 

	 \begin{equation}
	 {\mathbf 1_1} ~~:~~
	 x_1 y_2 + x_2 y_1, 
	 \end{equation}
	 and 

	 \begin{equation}
	 {\mathbf 1_2} ~~:~~
	 x_1 y_2 - x_2 y_1, 
	 \end{equation}
	 up to some normalization constant. From here on, we
	 follow the same method to obtain the vector spaces
	 that make up the Kronecker products, however we shall
	 neglect the work and list only the results. Lastly,
         for those representations that do depend on a parameter
	 (e.g. ${\mathbf 3_1}_{(l)}$) we determine the value of
	 the parameter by noting how the representations
	 transform under the action of $c$ and then finding
	 the corresponding values of the parameters as based
	 on Eq.~(\ref{eq:repsa}).

         \item[$\bullet$]{${\mathbf2_1} \otimes {\mathbf 3_1}_{(l)}={\mathbf
	  3_1}_{(l)}+{\mathbf 3_2}_{(l)}$.} Vector $x$ will
	  transform as a ${\mathbf 2_1}$ and $y$ as a ${\mathbf
	  3_1}_{(l)}$.

	 \begin{equation}
	 {\mathbf 3_1}_{(l)} ~~:~~
	 \begin{pmatrix}
	 x_1 y_1  \\ 
	 \omega x_1 y_2  \\ 
	 \omega^2 x_1 y_3  \\ 
	 \end{pmatrix}
	 +
	 \begin{pmatrix}
	 \omega^2 x_2 y_1 \\ 
	 \omega x_2 y_2 \\ 
	 x_2 y_3 \\ 
	 \end{pmatrix}, 
	 ~~~~~~~~ 
	 {\mathbf 3_2}_{(l)} ~~:~~
	 \begin{pmatrix}
	 x_1 y_1 \\ 
	 \omega x_1 y_2 \\ 
	 \omega^2 x_1 y_3\\ 
	 \end{pmatrix}
	 -
	 \begin{pmatrix}
	 \omega^2 x_2 y_1 \\ 
	 \omega x_2 y_2 \\ 
	 x_2 y_3 \\ 
	 \end{pmatrix}.
	 \end{equation}

         \item[$\bullet$]{${\mathbf2_1} \otimes {\mathbf 3_2}_{(l)}={\mathbf
	  3_1}_{(l)}+{\mathbf 3_2}_{(l)}$.} Vector $x$ will
	  transform as a ${\mathbf 2_1}$ and $y$ as a ${\mathbf
	  3_2}_{(l)}$.

	 \begin{equation}
	 {\mathbf 3_1}_{(l)} ~~:~~
	 \begin{pmatrix}
	 x_1 y_1  \\ 
	 \omega x_1 y_2 \\ 
	 \omega^2 x_1 y_3 \\ 
	 \end{pmatrix}
	 -
	 \begin{pmatrix}
	 \omega^2 x_2 y_1 \\ 
	 \omega x_2 y_2 \\ 
	 x_2 y_3 \\
	 \end{pmatrix}, 
	 ~~~~~~~~ 
	 {\mathbf 3_2}_{(l)} ~~:~~
	 \begin{pmatrix}
	 x_1 y_1  \\ 
	 \omega x_1 y_2  \\ 
	 \omega^2 x_1 y_3  \\ 
	 \end{pmatrix}
	 +
	 \begin{pmatrix}
	 \omega^2 x_2 y_1 \\ 
	 \omega x_2 y_2 \\ 
	 x_2 y_3 \\ 
	 \end{pmatrix}. 
	 \end{equation}

         \item[$\bullet$]{${\mathbf2_1} \otimes {\mathbf 6}_{(k,l)}={\mathbf
	  6}_{(k,l)}+{\mathbf 6}_{(k,l)}$.} Vector $x$ will
	  transform as a ${\mathbf 2_1}$ and $y$ as a ${\mathbf
	  6}_{(k,l)}$.

	 \begin{equation}
	 {\mathbf 6}_{(k,l)} ~~:~~
	 \begin{pmatrix}
	 x_1 y_1 \\ 
	 \omega x_1 y_2 \\ 
	 \omega^2 x_1 y_3 \\ 
	 x_2 y_4 \\ 
	 \omega x_2 y_5 \\ 
	 \omega^2 x_2 y_6 
	 \end{pmatrix}, 
	 ~~~~~~~~ 
	 {\mathbf 6}_{(k,l)} ~~:~~
	 \begin{pmatrix}
	 x_2 y_1 \\ 
	 \omega^2 x_2 y_2 \\ 
	 \omega x_2 y_3 \\ 
	 x_1 y_4 \\ 
	 \omega^2 x_1 y_5 \\ 
	 \omega x_1 y_6
	 \end{pmatrix}. 
	 \end{equation}

         \item[$\bullet$]{${\mathbf3_1}_{(l)} \otimes {\mathbf 3_1}_{(l')}={\mathbf
	  3_1}_{(l+l')}+{\mathbf 6}_{\widetilde{(l,-l')}}$.} 
	  Vector $x$ will transform as a ${\mathbf 3_1}_{(l)}$ and 
	  $y$ as a ${\mathbf 3_1}_{(l')}$.

	 \begin{equation}
	 {\mathbf 3_1}_{(l+l')} ~~:~~
	 \begin{pmatrix}
	 x_1 y_1 \\ 
	 x_2 y_2 \\ 
	 x_3 y_3 \\ 
	 \end{pmatrix}, 
	 ~~~~~~~~ 
	 {\mathbf 6}_{(-l,l-l')} ~~:~~
	 \begin{pmatrix}
	 x_1 y_2 \\ 
	 x_2 y_3 \\ 
	 x_3 y_1 \\ 
	 x_3 y_2 \\ 
	 x_2 y_1 \\ 
	 x_1 y_3
	 \end{pmatrix}, 
	 \end{equation}
	 where we note that $(l,-l')$ is related to $(-l,l-l')$
	 via the matrix transformation $M^p_s$. Recall  
	 we want to use the standard values explaining the tilde
	 used in the product.

         \item[$\bullet$]{${\mathbf3_1}_{(l)} \otimes {\mathbf 3_2}_{(l')}={\mathbf
	  3_2}_{(l+l')}+{\mathbf 6}_{\widetilde{(l,-l')}}$.} 
	  Vector $x$ will transform as a ${\mathbf 3_1}_{(l)}$ and 
	  $y$ as a ${\mathbf 3_2}_{(l')}$.

	 \begin{equation}
	 {\mathbf 3_2}_{(l+l')} ~~:~~
	 \begin{pmatrix}
	 x_1 y_1 \\ 
	 x_2 y_2 \\ 
	 x_3 y_3 \\ 
	 \end{pmatrix}, 
	 ~~~~~~~~ 
	 {\mathbf 6}_{(-l,l-l')} ~~:~~
	 \begin{pmatrix}
	 x_1 y_2 \\ 
	 x_2 y_3 \\ 
	 x_3 y_1 \\ 
	 -x_3 y_2 \\ 
	 -x_2 y_1 \\ 
	 -x_1 y_3
	 \end{pmatrix}, 
	 \end{equation}
	 where $(l,-l')$ is related to $(-l,l-l')$
	 via the matrix transformation $M^p_s$.

         \item[$\bullet$]{${\mathbf3_1}_{(l)} \otimes {\mathbf 6}_{(k',l')}=
	 {\mathbf
	 6}_{\widetilde{\tiny{\begin{pmatrix}k'\\l'-l\end{pmatrix}}}}
	 +
	 {\mathbf
	 6}_{\tiny{\widetilde{\begin{pmatrix}k'-l\\l'+l\end{pmatrix}}}}
	 +
	 {\mathbf
	 6}_{\tiny{\widetilde{\begin{pmatrix}l+k'\\l'\end{pmatrix}}}}$.}
	  Vector $x$ will transform as a ${\mathbf 3_1}_{(l)}$ and 
	  $y$ as a ${\mathbf 6}_{(k',l')}$.

	 \begin{equation}
         {\mathbf
	 6}_{\tiny{{\begin{pmatrix}l'-l\\l-k'-l'\end{pmatrix}}}}
	 ~~:~~
	 \begin{pmatrix}
	 x_1 y_3 \\ 
	 x_2 y_1 \\ 
	 x_3 y_2 \\ 
	 x_3 y_6 \\ 
	 x_2 y_4 \\ 
	 x_1 y_5
	 \end{pmatrix}, 
	 ~~~~~~~~ 
	 {\mathbf
	 6}_{{\tiny{\begin{pmatrix}k'-l\\l'+l\end{pmatrix}}}}
	 ~~:~~
	 \begin{pmatrix}
	 x_1 y_1 \\ 
	 x_2 y_2 \\ 
	 x_3 y_3 \\ 
	 x_3 y_4 \\ 
	 x_2 y_5 \\ 
	 x_1 y_6 \\ 
	 \end{pmatrix}, 
	 ~~~~~~~~ 
         {\mathbf
	 6}_{\tiny{{\begin{pmatrix}-l-k'-l'\\l+k'\end{pmatrix}}}}
	 ~~:~~
	 \begin{pmatrix}
	 x_1 y_2 \\ 
	 x_2 y_3 \\ 
	 x_3 y_1 \\ 
	 x_3 y_5 \\ 
	 x_2 y_6 \\ 
	 x_1 y_4 \\ 
	 \end{pmatrix},
         \end{equation}
	 where $(l'-l,l-k'-l')$ is related to $(k',l'-l)$ by
	 the matrix transformation $M^p_s$, and similarly 
	 $(-l-k'-l',l+k')$ is related to $(l+k',l')$.

         \item[$\bullet$]{${\mathbf3_2}_{(l)} \otimes {\mathbf 3_2}_{(l')}={\mathbf
	  3_1}_{(l+l')}+{\mathbf 6}_{\widetilde{(l,-l')}}$.} 
	  Vector $x$ will transform as a ${\mathbf 3_2}_{(l)}$ and 
	  $y$ as a ${\mathbf 3_2}_{(l')}$.

	 \begin{equation}
	 {\mathbf 3_1}_{(l+l')} ~~:~~
	 \begin{pmatrix}
	 x_1 y_1 \\ 
	 x_2 y_2 \\ 
	 x_3 y_3 \\ 
	 \end{pmatrix}, 
	 ~~~~~~~~ 
	 {\mathbf 6}_{(-l,l-l')} ~~:~~
	 \begin{pmatrix}
	 x_1 y_2 \\ 
	 x_2 y_3 \\ 
	 x_3 y_1 \\ 
	 x_3 y_2 \\ 
	 x_2 y_1 \\ 
	 x_1 y_3
	 \end{pmatrix}, 
	 \end{equation}
	 where we note that $(l,-l')$ is related to $(-l,l-l')$
	 via the matrix transformation $M^p_s$. 

         \item[$\bullet$]{${\mathbf3_2}_{(l)} \otimes {\mathbf 6}_{(k',l')}
	 =
	 {\mathbf
	 6}_{\widetilde{\tiny{\begin{pmatrix}k'\\l'-l\end{pmatrix}}}}
	 +
	 {\mathbf
	 6}_{\tiny{\widetilde{\begin{pmatrix}k'-l\\l'+l\end{pmatrix}}}}
	 +
	 {\mathbf
	 6}_{\tiny{\widetilde{\begin{pmatrix}l+k'\\l'\end{pmatrix}}}}$.}
	  Vector $x$ will transform as a ${\mathbf 3_2}_{(l)}$ and 
	  $y$ as a ${\mathbf 6}_{(k',l')}$.

	 \begin{equation}
         {\mathbf
	 6}_{\tiny{{\begin{pmatrix}l'-l\\l-k'-l'\end{pmatrix}}}}
	 ~~:~~
	 \begin{pmatrix}
	 x_1 y_3 \\ 
	 x_2 y_1 \\ 
	 x_3 y_2 \\ 
	 -x_3 y_6 \\ 
	 -x_2 y_4 \\ 
	 -x_1 y_5
	 \end{pmatrix}, 
	 ~~~~~~~~ 
	 {\mathbf
	 6}_{{\tiny{\begin{pmatrix}k'-l\\l'+l\end{pmatrix}}}}
	 ~~:~~
	 \begin{pmatrix}
	 x_1 y_1 \\ 
	 x_2 y_2 \\ 
	 x_3 y_3 \\ 
	 -x_3 y_4 \\ 
	 -x_2 y_5 \\ 
	 -x_1 y_6 \\ 
	 \end{pmatrix}, 
	 ~~~~~~~~ 
         {\mathbf
	 6}_{\tiny{{\begin{pmatrix}-l-k'-l'\\l+k'\end{pmatrix}}}}
	 ~~:~~
	 \begin{pmatrix}
	 x_1 y_2 \\ 
	 x_2 y_3 \\ 
	 x_3 y_1 \\ 
	 -x_3 y_5 \\ 
	 -x_2 y_6 \\ 
	 -x_1 y_4 \\ 
	 \end{pmatrix},
         \end{equation}
	 where $(l'-l,l-k'-l')$ is related to $(k',l'-l)$ by
	 the matrix transformation $M^p_s$, and similarly 
	 $(-l-k'-l',l+k')$ is related to $(l+k',l')$.

         \item[$\bullet$]{${\mathbf6}_{(k,l)} \otimes {\mathbf 6}_{(k',l')}=\sum_{p,s}
	 {\mathbf
	 6}_{\widetilde{\tiny{\left (
	 \begin{pmatrix}k\\l\end{pmatrix}
	 +
	 M^p_s
	 \begin{pmatrix}k'\\l'\end{pmatrix}\right )}}}$}
	  Vector $x$ will transform as a ${\mathbf 6}_{(k,l)}$ and 
	  $y$ as a ${\mathbf 6}_{(k',l')}$.

	 \begin{equation}
         {\mathbf
	 6}_{\tiny{{\begin{pmatrix}k+k'\\l+l'\end{pmatrix}}}}
	 ~~:~~
	 \begin{pmatrix}
	 x_1 y_1 \\ 
	 x_2 y_2 \\ 
	 x_3 y_3 \\ 
	 x_4 y_4 \\ 
	 x_5 y_5 \\ 
	 x_6 y_6
	 \end{pmatrix}, 
	 ~~~~~~~~ 
	 {\mathbf
	 6}_{{\tiny{\begin{pmatrix}k-k'-l'\\l+k'\end{pmatrix}}}}
	 ~~:~~
	 \begin{pmatrix}
	 x_1 y_2 \\ 
	 x_2 y_3 \\ 
	 x_3 y_1 \\ 
	 x_4 y_5 \\ 
	 x_5 y_6 \\ 
	 x_6 y_4 \\ 
	 \end{pmatrix}, 
	 ~~~~~~~~ 
         {\mathbf
	 6}_{\tiny{{\begin{pmatrix}k+l'\\l-l'-k'\end{pmatrix}}}}
	 ~~:~~
	 \begin{pmatrix}
	 x_1 y_3 \\ 
	 x_2 y_1 \\ 
	 x_3 y_2 \\ 
	 x_4 y_6 \\ 
	 x_5 y_4 \\ 
	 x_6 y_5 \\ 
	 \end{pmatrix}, \nonumber
         \end{equation}

	 \begin{equation}
         {\mathbf
	 6}_{\tiny{{\begin{pmatrix}k-k'\\l+k'+l'\end{pmatrix}}}}
	 ~~:~~
	 \begin{pmatrix}
	 x_1 y_4 \\ 
	 x_2 y_6 \\ 
	 x_3 y_5 \\ 
	 x_4 y_1 \\ 
	 x_5 y_3 \\ 
	 x_6 y_2
	 \end{pmatrix}, 
	 ~~~~~~~~ 
	 {\mathbf
	 6}_{{\tiny{\begin{pmatrix}k+k'+l'\\l-l'\end{pmatrix}}}}
	 ~~:~~
	 \begin{pmatrix}
	 x_1 y_5 \\ 
	 x_2 y_4 \\ 
	 x_3 y_6 \\ 
	 x_4 y_2 \\ 
	 x_5 y_1 \\ 
	 x_6 y_3 \\ 
	 \end{pmatrix}, 
	 ~~~~~~~~ 
         {\mathbf
	 6}_{\tiny{{\begin{pmatrix}k-l'\\l-k'\end{pmatrix}}}}
	 ~~:~~
	 \begin{pmatrix}
	 x_1 y_6 \\ 
	 x_2 y_5 \\ 
	 x_3 y_4 \\ 
	 x_4 y_3 \\ 
	 x_5 y_2 \\ 
	 x_6 y_1 \\ 
	 \end{pmatrix}.
         \end{equation}

         \end{itemize}

     \item[$(ii)$] {$n=3\mathbb Z$.} The only major difference
     between the case here and case~($i$) is the addition of three
     more two-dimensional representations. Here, we limit ourselves to all products which 
     involve the irreducible representation ${\bf 2_2}$. The products
     involving ${\bf 2_3}$ and ${\bf 2_4}$ are similar in structure and can be
     determined by the reader following similar lines. To begin, the 
     additional two-dimensional representations transform under
     $a$, $b$, and $c$ as: 

     \begin{eqnarray}
     {\mathbf 2_2} &:&~~~~
     \begin{pmatrix} x_1 \\ x_2 
     \end{pmatrix}
     \mapsto
     \begin{pmatrix} \omega x_1 \\ \omega^2 x_2 
     \end{pmatrix}_{a}
     ,~~~
     \begin{pmatrix} x_2 \\ x_1 
     \end{pmatrix}_{b}
     ,~~~
     \begin{pmatrix} \omega^2 x_1 \\ \omega x_2 
     \end{pmatrix}_{c}, \notag
     \\
     {\mathbf 2_3} &:&~~~~
     \begin{pmatrix} x_1 \\ x_2 
     \end{pmatrix}
     \mapsto
     \begin{pmatrix} \omega x_1 \\ \omega^2 x_2 
     \end{pmatrix}_{a}
     ,~~~
     \begin{pmatrix} x_2 \\ x_1 
     \end{pmatrix}_{b}
     ,~~~
     \begin{pmatrix} \omega x_1 \\ \omega ^2 x_2 
     \end{pmatrix}_{c},
     \\
     {\mathbf 2_4} &:&~~~~
     \begin{pmatrix} x_1 \\ x_2 
     \end{pmatrix}
     \mapsto
     \begin{pmatrix} x_1 \\ x_2 
     \end{pmatrix}_{a}
     ,~~~
     \begin{pmatrix} x_2 \\ x_1 
     \end{pmatrix}_{b}
     ,~~~
     \begin{pmatrix} \omega x_1 \\ \omega^2 x_2 
     \end{pmatrix}_{c}. \notag
     \end{eqnarray}

         \begin{itemize}
         
  \item[$\bullet$]{${\mathbf2_2} \otimes {\mathbf 2_1}={\mathbf
	  2_3}+ {\mathbf 2_4}$.} The quantity $x$ transforms as
	  a $\mathbf 2_2$ and $y$ as a $\mathbf 2_1$.

	 \begin{equation}
	 {\mathbf 2_3} ~~:~~
	 \begin{pmatrix}
	 x_2 y_2 \\ x_1 y_1
	 \end{pmatrix}, 
	 ~~~~~~~~ 
	 {\mathbf 2_4} ~~:~~
	 \begin{pmatrix}
	 x_2 y_1 \\ x_1 y_2.
	 \end{pmatrix}. 
	 \end{equation}

          \item[$\bullet$]{${\mathbf2_r} \otimes {\mathbf 2_r}={\mathbf
	  1_1}+ {\mathbf 1_2} +{\mathbf 2_r}$.} We can
	  generalize since the result is true regardless of
	  the choice of $r=1,2,3,4$. The quantity $x$ and $y$
	  both transform as ${\mathbf 2_r }.$

	 \begin{equation}
	 {\mathbf 2_r} ~~:~~
	 \begin{pmatrix}
	 x_2 y_2 \\ x_1 y_1
	 \end{pmatrix}. 
	 ~~~~~~~~ 
	 {\mathbf 1_1} ~~:~~
	 x_1 y_2 + x_2 y_1, 
	 ~~~~~~~~ 
	 {\mathbf 1_2} ~~:~~
	 x_1 y_2 - x_2 y_1, 
	 \end{equation}

          \item[$\bullet$]{${\mathbf2_2} \otimes {\mathbf 2_3}={\mathbf
	  2_1}+ {\mathbf 2_4}$.} The quantity $x$ transforms as
	  a $\mathbf 2_2$ and $y$ as a $\mathbf 2_3$.

	 \begin{equation}
	 {\mathbf 2_1} ~~:~~
	 \begin{pmatrix}
	 x_2 y_2 \\ x_1 y_1
	 \end{pmatrix}, 
	 ~~~~~~~~ 
	 {\mathbf 2_4} ~~:~~
	 \begin{pmatrix}
	 x_1 y_2 \\ x_2 y_1.
	 \end{pmatrix}. 
	 \end{equation}
  
  \item[$\bullet$]{${\mathbf2_2} \otimes {\mathbf 2_4}={\mathbf
	  2_1}+ {\mathbf 2_3}$.} The quantity $x$ transforms as
	  a $\mathbf 2_2$ and $y$ as a $\mathbf 2_4$.

	 \begin{equation}
	 {\mathbf 2_1} ~~:~~
	 \begin{pmatrix}
	 x_1 y_1 \\ x_2 y_2
	 \end{pmatrix}, 
	 ~~~~~~~~ 
	 {\mathbf 2_3} ~~:~~
	 \begin{pmatrix}
	 x_1 y_2 \\ x_2 y_1.
	 \end{pmatrix}. 
	 \end{equation}

          \item[$\bullet$]{${\mathbf2_2} \otimes {\mathbf
	  3_1}_{(l)}=
	  {\mathbf
	  6}_{\widetilde{\tiny{\begin{pmatrix}2n/3-l\\2n/3
	  +l\end{pmatrix}}}}$.} The quantity $x$ transforms as
	  a $\mathbf 2_2$ and $y$ as a ${\mathbf 3_1}_{(l)}$.

	 \begin{equation}
	  {\mathbf
	  6}_{{\tiny{\begin{pmatrix}2n/3-l\\2n/3
	  +l\end{pmatrix}}}} ~~:~~
	 \begin{pmatrix}
	 x_1 y_1 \\ 
	 \omega x_1 y_2 \\ 
	 \omega^2 x_1 y_3 \\ 
	 x_2 y_3 \\ 
	 \omega x_2 y_2 \\ 
	 \omega^2 x_2 y_1 \\ 
	 \end{pmatrix}. 
	 \end{equation}

          \item[$\bullet$]{${\mathbf2_2} \otimes {\mathbf
	  3_2}_{(l)}=
	  {\mathbf
	  6}_{\widetilde{\tiny{\begin{pmatrix}2n/3-l\\2n/3
	  +l\end{pmatrix}}}}$.} The quantity $x$ transforms as
	  a $\mathbf 2_2$ and $y$ as a ${\mathbf 3_2}_{(l)}$.

	 \begin{equation}
	  {\mathbf
	  6}_{{\tiny{\begin{pmatrix}2n/3-l\\2n/3
	  +l\end{pmatrix}}}} ~~:~~
	 \begin{pmatrix}
	 x_1 y_1 \\ 
	 \omega x_1 y_2 \\ 
	 \omega^2 x_1 y_3 \\ 
	 -x_2 y_3 \\ 
	 -\omega x_2 y_2 \\ 
	 -\omega^2 x_2 y_1 \\ 
	 \end{pmatrix}. 
	 \end{equation}

          \item[$\bullet$]{${\mathbf2_2} \otimes {\mathbf
	  6}_{(k,l)}=
	  {\mathbf
	  6}_{\widetilde{\tiny{\begin{pmatrix}k+n/3\\l+n/3
	  \end{pmatrix}}}} 
	  +
	  {\mathbf
	  6}_{\widetilde{\tiny{\begin{pmatrix}k+2n/3\\l+2n/3
	  \end{pmatrix}}}}$.} 
	  The quantity $x$ transforms as
	  a $\mathbf 2_2$ and $y$ as a ${\mathbf 6}_{(k,l)}$.

	 \begin{equation}
	  {\mathbf
	  6}_{{\tiny{\begin{pmatrix}k+n/3\\l+n/3
	  \end{pmatrix}}}} ~~:~~
	 \begin{pmatrix}
	 x_2 y_1 \\ 
	 \omega^2 x_2 y_2 \\ 
	 \omega x_2 y_3 \\ 
	 x_1 y_4 \\ 
	 \omega^2 x_1 y_5 \\ 
	 \omega x_1 y_6 \\ 
	 \end{pmatrix}, 
	 ~~~~~~~~ 
	  {\mathbf
	  6}_{{\tiny{\begin{pmatrix}k+2n/3\\l+2n/3
	  \end{pmatrix}}}} ~~:~~
	 \begin{pmatrix}
	 x_1 y_1 \\ 
	 \omega x_1 y_2 \\ 
	 \omega^2 x_1 y_3 \\ 
	 x_2 y_4 \\ 
	 \omega x_2 y_5 \\ 
	 \omega^2 x_2 y_6 \\ 
	 \end{pmatrix}. 
	 \end{equation}

         \end{itemize}	
     \end{itemize}

\end{itemize}



\begin{thebibliography}{99}



\bibitem{tribi}
  P.~F. Harrison, D.~H. Perkins, and W.~G. Scott, {\em Phys. Lett.} B530:167,
  2002, {hep-ph/0202074};\\
  P.~F. Harrison and W.~G. Scott, {\em Phys. Lett.} B535:163, 2002,
  {hep-ph/0203209}. 

\bibitem{S3}
  S.~Pakvasa and H.~Sugawara, {\em Phys. Lett.} B73:61, 1978; \\
  J.~Kubo, A.~Mondragon, M.~Mondragon and E.~Rodriguez-Jauregui, {\em Prog.
    Theor. Phys.} 109:795, 2003, Erratum: {\em ibid.} 114:287, 2005,
    {hep-ph/0302196};\\
  F.~Caravaglios and S.~Morisi, 2005, {hep-ph/0503234};\\
  W.~Grimus and L.~Lavoura, {\em JHEP} 0508:013, 2005, {hep-ph/0504153};\\
  W.~Grimus and L.~Lavoura, {\em JHEP} 0601:018, 2006, {hep-ph/0509239};\\
  R.~N. Mohapatra, S.~Nasri, and H.-B. Yu, {\em Phys. Lett.} B639:318, 2006,
    {hep-ph/0605020};  \\
  R.~Jora, S.~Nasri, and J.~Schechter, {\em Int. J. Mod. Phys.} A21:5875,
    2006, {hep-ph/0605069};\\
  O.~Felix, A.~Mondragon, M.~Mondragon and E.~Peinado, {\em Rev. Mex. Fis.}
     S52:67, 2006, {hep-ph/0610061};\\
  Y.~Koide, {\em Eur. Phys. J.} C50:809, 2007, {hep-ph/0612058};\\
  K.~S.~Babu, A.~G.~Bachri and Z.~Tavartkiladze, {\em Int. J. Mod. Phys.}
  A23:1679, 2008, {arXiv:0705.4419};\\
  F.~Feruglio and Y.~Lin, {\em Nucl. Phys.}  B800:77, 2008, {arXiv:0712.1528}.

\bibitem{A4}
  E.~Ma and G.~Rajasekaran, {\em Phys. Rev.} D64:113012, 2001,
    {hep-ph/0106291}; \\
  K.~S. Babu, E.~Ma, and J.~W.~F. Valle, {\em Phys. Lett.} B552:207, 2003,
    {hep-ph/0206292}; \\
  G.~Altarelli and F.~Feruglio, {\em Nucl. Phys.} B720:64, 2005,
    {hep-ph/0504165};\\
  K.~S. Babu and X.-G. He, 2005, {hep-ph/0507217};\\
  A.~Zee, {\em Phys. Lett.} B630:58, 2005, {hep-ph/0508278};\\
  G.~Altarelli and F.~Feruglio, {\em Nucl. Phys.} B741:215, 2006,
    {hep-ph/0512103};\\
  E.~Ma, {\em Mod. Phys. Lett.}  A21:2931, 2006, {arXiv:hep-ph/0607190};\\
  S.~F. King and M.~Malinsky, {\em Phys. Lett.} B645:351, 2007,
    {hep-ph/0610250};\\
  S.~Morisi, M.~Picariello, and E.~Torrente-Lujan, {\em Phys. Rev.}
    D75:075015, 2007, {hep-ph/0702034};\\
  F.~Feruglio, C.~Hagedorn, Y.~Lin, and L.~Merlo,  {\em Nucl. Phys.}
    B775:120, 2007, {hep-ph/0702194}; \\
  M.~Hirsch, A.~S. Joshipura, S.~Kaneko, and J.~W.~F. Valle, {\em
    Phys. Rev. Lett.} 99:151802, 2007, {hep-ph/0703046};\\
  F.~Bazzocchi, S.~Morisi and M.~Picariello, {\em Phys. Lett.}  B659:628, 2008,
  {arXiv:0710.2928};\\
  M.~Honda and M.~Tanimoto, {\em Prog. Theor. Phys.}  119:583, 2008,
  {arXiv:0801.0181};\\
  G.~Altarelli, F.~Feruglio and C.~Hagedorn, {\em  JHEP} 0803:052, 2008,
  {arXiv:0802.0090};\\
  Y.~Lin, 2008, {arXiv:0804.2867}.

\bibitem{S4}
  C.~Hagedorn, M.~Lindner, and R.~N. Mohapatra, {\em JHEP} 0606:042, 2006,
   {hep-ph/0602244};\\
  Y.~Koide, {\em JHEP} 0708:086, 2007, {arXiv:0705.2275};\\
  C.~S.~Lam, 2008, {arXiv:0804.2622}.

\bibitem{delta27}
  I.~de~Medeiros~Varzielas and G.~G. Ross, {\em Nucl. Phys.} B733:31, 2006,
    {hep-ph/0507176};\\
  I.~de~Medeiros~Varzielas, S.~F. King, and G.~G. Ross, {\em Phys. Lett.}
    B644:153, 2007, {hep-ph/0512313};\\
  I.~de~Medeiros~Varzielas, S.~F. King, and G.~G. Ross, {\em Phys. Lett.}
    B648:201, 2007, {hep-ph/0607045};\\
  E.~Ma, {\em Mod. Phys. Lett.} A21:1917, 2006, {hep-ph/0607056};\\
  E.~Ma, {\em Phys. Lett.} B660:505, 2008, {arXiv:0709.0507}. 

\bibitem{binaryA4}
  A.~Aranda, C.~D.~Carone and R.~F.~Lebed, {\em Phys. Rev.}  D62:016009, 2000,
  {hep-ph/0002044};\\
  F.~Feruglio, C.~Hagedorn, Y.~Lin and L.~Merlo, {\em Nucl. Phys.}  B775:120,
  2007, {hep-ph/0702194};\\
  M.~C.~Chen and K.~T.~Mahanthappa, {\em Phys. Lett.}  B652:34, 2007,
    {arXiv:0705.0714};\\
  P.~H.~Frampton and T.~W.~Kephart, {\em  JHEP} 0709:110, 2007,
  {arXiv:0706.1186};\\
  A.~Aranda, {\em  Phys. Rev.}  D76:111301, 2007,
  {arXiv:0707.3661}.

\bibitem{Q2n}
  D.~Chang, W.~Y.~Keung and G.~Senjanovic, {\em Phys. Rev.}  D42:1599, 1990;\\
  D.~Chang, W.~Y.~Keung, S.~Lipovaca and G.~Senjanovic, {\em Phys. Rev. Lett.}
  67:953, 1991;\\ 
  P.~H.~Frampton and T.~W.~Kephart, {\em Phys. Rev.}  D51:1, 1995,
  {hep-ph/9409324};\\
  P.~H.~Frampton and T.~W.~Kephart, {\em Int. J. Mod. Phys.}  A10:4689, 1995,
  {hep-ph/9409330};\\
  P.~H.~Frampton and O.~C.~W.~Kong, {\em Phys. Rev. Lett.}  75:781, 1995,
  {hep-ph/9502395};\\
  P.~H.~Frampton and O.~C.~W.~Kong, {\em Phys. Rev.}  D53:2293, 1996,
  {hep-ph/9511343};\\
  P.~H.~Frampton and O.~C.~W.~Kong, {\em Phys. Rev. Lett.}  77:1699, 1996,
  {hep-ph/9603372};\\
  M.~Frigerio, S.~Kaneko, E.~Ma and M.~Tanimoto, {\em Phys. Rev.}  D71:011901,
  2005, {hep-ph/0409187};\\
  K.~S.~Babu and J.~Kubo, {\em Phys. Rev.}  D71:056006, 2005,
  {hep-ph/0411226};\\
  M.~Frigerio, 2005, {hep-ph/0505144};\\
  Y.~Kajiyama, E.~Itou and J.~Kubo, {\em  Nucl. Phys.}  B743:74, 2006,
  {hep-ph/0511268};\\
  M.~Frigerio and E.~Ma, {\em Phys. Rev.}  D76:096007, 2007,
  {arXiv:0708.0166}.

\bibitem{D5}
  C.~Hagedorn, M.~Lindner, and F.~Plentinger, {\em Phys. Rev.} D74:025007,
  2006, {hep-ph/0604265};\\
  A.~Blum, C.~Hagedorn, and M.~Lindner, {\em Phys. Rev.} D77:076004, 2008,
  {arXiv:0709.3450}. 

\bibitem{psl}
  C.~Luhn, S.~Nasri and P.~Ramond, {\em Phys. Lett.}  B652:27, 2007,
  {arXiv:0706.2341};\\
  C.~Luhn, S.~Nasri and P.~Ramond, {\em J. Math. Phys.}  48:123519, 2007,
  {arXiv:0709.1447}.

\bibitem{seidl}
  F.~Plentinger, G.~Seidl and W.~Winter, {\em JHEP} 0804:077, 2008,
  {arXiv:0802.1718};\\
  F.~Plentinger and G.~Seidl, {\em Phys. Rev.}  D78:045004, 2008,
  {arXiv:0803.2889}.

\bibitem{other}
  D.~B. Kaplan and M.~Schmaltz, {\em Phys. Rev.} D49:3741, 1994,
    {hep-ph/9311281};\\
  M.~Schmaltz, {\em Phys. Rev.} D52:1643, 1995, {hep-ph/9411383};\\
  E.~Ma, {\em Europhys. Lett.} 79:61001, 2007,
  {hep-ph/0701016}.

\bibitem{overview}
  E.~Ma, 2007, {arXiv:0705.0327};\\
  G.~Altarelli, 2007, {arXiv:0705.0860};\\
  C.~S.~Lam, {\em Phys. Lett.}  B656:193, 2007,
  {arXiv:0708.3665}.

\bibitem{finitesubgroup}
  G.~A. Miller, H.~F. Blichfeldt, and L.~E. Dickson,
  {\em Theory and Applications of Finite Groups}, John Wiley \& Sons,
  New York 1916, and Dover Edition 1961;\\
  W.~M. Fairbairn, T.~Fulton, and W.~H. Klink, {\em J. Math. Phys.} 5:1038,
  1964. 

\bibitem{LNR}
  C.~Luhn, S.~Nasri, and P.~Ramond, {\em J. Math. Phys.}  48:073501, 2007,
    {hep-th/0701188}.

\bibitem{Bovier:1980gc}
A.~Bovier, M.~L{\"u}ling, and D.~Wyler.
\newblock {\em J. Math. Phys.}, 22:1543, 1981.









\end{thebibliography}
\end{document}